\documentclass[preprint]{aastex}

\begin{document}

\title{OPTICAL PHOTOMETRY OF THE TYPE Ia SN~1999ee AND THE TYPE Ib/c SN~1999ex IN IC 5179}
\author{Maximilian Stritzinger\altaffilmark{1}}
\affil{Department of Physics, The University of Arizona, Tucson, AZ 85721}
\email{stritzin@as.arizona.edu}
\author{Mario Hamuy\altaffilmark{1,2}}
\affil{The Observatories of the Carnegie Institution of Washington, 813 Santa Barbara Street, Pasadena, CA 91101}
\email{mhamuy@ociw.edu}      
\author{Nicholas B. Suntzeff}
\affil{National Optical Astronomy Observatories\altaffilmark{3}, Cerro Tololo Inter-American Observatory, Casilla 603, La Serena, Chile}
\email{nsuntzeff@noao.edu}
\author{R. C. Smith}
\affil{National Optical Astronomy Observatories\altaffilmark{3}, Cerro Tololo Inter-American Observatory, Casilla 603, La Serena, Chile}
\email{csmith@noao.edu}
\author{M. M. Phillips}
\affil{Carnegie Institution of Washington, Las Campanas Observatory, Casilla 601, La Serena, Chile}
\email{mmp@lco.cl}
\author{Jos\'{e} Maza}
\affil{Departamento de Astronom\'\i a, Universidad de Chile, Casilla 36-D, Santiago, Chile}
\email {jose@das.uchile.cl}    
\author{L. -G. Strolger}
\affil{Department of Astronomy, University of Michigan, Ann Arbor, MI 48109-1090}
\email{loust@umich.edu}
\author{Roberto Antezana}
\affil{Departamento de Astronom\'\i a, Universidad de Chile, Casilla 36-D, Santiago, Chile}
\email {rantezan@das.uchile.cl}
\author{Luis Gonz\'alez} 
\affil{Departamento de Astronom\'\i a, Universidad de Chile, Casilla 36-D, Santiago, Chile}
\email {lgonzale@das.uchile.cl}
\author{Marina Wischnjewsky}
\affil{Departamento de Astronom\'\i a, Universidad de Chile, Casilla 36-D, Santiago, Chile}
\email {marina@das.uchile.cl}
\author{Pablo Candia}
\affil{National Optical Astronomy Observatories\altaffilmark{3}, Cerro Tololo Inter-American Observatory, Casilla 603, La Serena, Chile}
\email{candia@noao.edu}
\author{Juan Espinoza}
\affil{National Optical Astronomy Observatories\altaffilmark{3}, Cerro Tololo Inter-American Observatory, Casilla 603, La Serena, Chile}
\author{David Gonz\'alez}
\affil{National Optical Astronomy Observatories\altaffilmark{3}, Cerro Tololo Inter-American Observatory, Casilla 603, La Serena, Chile}
\author{Christopher Stubbs}
\affil{Department of Astronomy, University of Washington, Box 351580, Seattle, WA 98195-1580}
\email{stubbs@astro.washington.edu}
\author{A. C. Becker\altaffilmark{4}}
\affil{Departments of Astronomy and Physics, University of Washington, Seattle, WA 98195}
\email{becker@astro.washington.edu}
\author{Eric P. Rubenstein}
\affil{Department of Astronomy, Yale University, P.O. Box 208101, New Haven, CT 06520-8101}
\email{ericr@astro.yale.edu}
\author{Gaspar Galaz\altaffilmark{5}}
\affil{Carnegie Institution of Washington, Las Campanas Observatory, Casilla 601, La Serena, Chile}
\email{gaspar@azul.lco.cl}

\altaffiltext{1}{Visiting Astronomer, Cerro Tololo Inter-American Observatory.
CTIO is operated by AURA, Inc.\ under contract to the National Science
Foundation.}
\altaffiltext{2}{Hubble Fellow.}
\altaffiltext{3}{Cerro Tololo Inter-American Observatory, Kitt Peak
National Observatory, National Optical Astronomy Observatories,
operated by the Association of Universities for Research in Astronomy,
Inc., (AURA), under cooperative agreement with the National Science
Foundation.}
\altaffiltext{4}{Present address: Bell Laboratories, Lucent Technologies, 600 Mountain Avenue, Murray Hill, NJ 07974.}
\altaffiltext{5}{Present address: Departamento de Astronom\'\i a y Astrof\'\i sica, P. Universidad Cat\'olica de Chile, Casilla 306, Santiago, Chile}

\begin{abstract}
We present $UBVRIz$ lightcurves of the Type~Ia SN~1999ee and the Type~Ib/c 
SN~1999ex, both located in the galaxy IC~5179. SN~1999ee has an extremely
well sampled lightcurve spanning from 10 days before $B_{max}$ through 53
days after peak. Near maximum we find systematic differences $\sim$0.05 mag
in photometry measured with two different telescopes, even though
the photometry is reduced to the same local standards around the supernova
using the specific color terms for each instrumental system. We use
models for our bandpasses and spectrophotometry of SN~1999ee
to derive magnitude corrections (S-corrections) and remedy this problem. 
This exercise demonstrates the need of accurately characterizing the instrumental
system before great photometric accuracies of Type Ia supernovae can be claimed. It also shows
that this effect can have important astrophysical consequences since a small
systematic shift of 0.02 mag in the $B-V$ color can introduce a $0.08$ mag error
in the extinction corrected peak $B$ magnitudes of a supernova and thus
lead to biased cosmological parameters. The data for the Type~Ib/c
SN~1999ex present us with the first ever observed shock breakout of a 
supernova of this class. These observations show that shock breakout occurred
18 days before $B_{max}$ and support the idea that Type~Ib/c supernovae are due
to core collapse of massive stars rather than thermonuclear disruption of white dwarfs.

\end{abstract}

\keywords{supernovae: Type~Ia, Type~Ib/c --optical photometry --SN~1999ee --SN~1999ex}

\section{Introduction}

In recent years significant effort has gone into searching for and observing Type~Ia 
supernovae (hereafter referred to as SNe) with the purpose of determining cosmological parameters.
Despite this progress, most of SNe~Ia observations have been limited to optical wavelengths, and
relatively little is still known about the infrared (IR) properties of these objects. IR photometry
of SNe has been limited to  a handful of events, including the early work of \citet{elias81,elias85}
and \citet{frogel87}, and the most recent efforts by \citet{jha99}, \citet{hernandez00}, and \citet{krisciunas00}.
Combined IR studies of SNe~Ia \citep{meikle00} have shown that SNe~Ia display a scatter in peak 
absolute magnitude no larger than $\pm0.15$ mag in all three IR bands $JHK$,
which is comparable to that obtained in the optical. To further exploit the usefulness of
these objects as distance indicators, as well as to better understand the explosion mechanism
and nature of SNe, more IR observations are clearly required.

Today the advent of new IR detectors enables us to obtain high quality observations
and enlarge the hitherto small samples of SNe observed at these wavelengths.
The ``Supernova Optical and Infrared Survey'' (SOIRS) was initiated in 1999 to make extensive IR and
optical observations of bright candidates from the El Roble 
SN survey \citep{maza81}, and from the Nearby Galaxies Supernova Search \citep{strolger99}.
Once objects were identified to be SN candidates, followup observations were
conducted at several observatories including the Cerro Tololo Inter-American Observatory
(CTIO) for optical/IR photometry, the Las Campanas Observatory (LCO) for IR photometry,
and the European Southern Observatory (ESO) at Cerro Paranal and La Silla for optical/IR
spectroscopy. During 1999-2000 this campaign resulted in a detailed study of 18 SNe.
Of these events 10 were SNe~Ia, 7 SNe~II and 1 a Type Ib/c event.  

This paper presents optical photometry for the two best-observed  SNe
in this survey, SN~1999ee and SN~1999ex. SN~1999ee was discovered by
M. Wischnjewsky on a T-Max 400 film obtained by L. Gonz\'alez on 1999 October 7.15 UT (JD 2,451,458.65)
in the course of the El Roble survey \citep{maza99}. This SN exploded in IC~5179, a very active star-forming
Sc galaxy with a heliocentric redshift of 3,498 $km$ $s^{-1}$.
Adopting a value for $H_{\circ}$=63.3$\pm$3.5 $km$ $s^{-1}$ $Mpc^{-1}$ \citep{phillips99}
along with a redshift of 3,239$\pm$300 $km$ $s^{-1}$ in the Cosmic
Microwave Background (CMB) Reference System, IC~5179 is located at
a distance of 51.2$\pm$5.5 Mpc ($\mu$=33.55$\pm$0.23),
which we adopt throughout this paper.
An optical spectrum taken on 1999 October 9.10 revealed that SN~1999ee was
a very young SN~Ia, which led us to give it a high priority among our list of
targets for followup observations.

In a rare occurrence IC~5179 produced a second SN
within a few weeks from the discovery of SN~1999ee. The second SN (1999ex)
was discovered on 1999 November 9.51 UT by \cite{martin99}
at Perth Observatory through the course of the PARG Automated
Supernova Search. Initially we classified it as a Ic event \citep{hamuy99}
based on the close resemblance to SN~1994I \citep{filippenko95}, although
we now believe it belongs to an intermediate Ib/c class \citep{hamuy02}.
Although SN~1999ex was discovered on 1999 November 9,
examination of our data revealed that the object was present on
our CCD images obtained 10 days before discovery.
Further examination of the data showed that we had unintentionally
detected the initial shock breakout resulting from the core bounce of SN~1999ex which,
until now, has never been observed in a Ib/c event. 

The observations gathered for SNe~1999ee and 1999ex have an unprecedented
temporal and  wavelength coverage and afford the opportunity to carry out a detailed comparison
with explosion and atmosphere models. In this paper we report final reduced $UBVRIz$ lightcurves for both 
of these SNe. The photometry presented here is available in electronic form
to other researchers.\footnote{At http://www.ociw.edu/$\sim$mhamuy/lightcurves.html.}
Section \ref{obs.sec} discusses the observations and the data
reduction process and Section \ref{results.sec} presents final results for both events.
We proceed with a discussion in Section \ref{disc.sec} after which we summarize
our conclusions.  Readers are to refer to K. Krisciunas et al. (2002, in preparation) and \citet{hamuy02}, which include
the analysis of IR photometry and optical/IR spectra, respectively.

\section{Observations}
\label{obs.sec}

\begin{figure}
\figurenum{1}
\epsscale{1.0}
\plotone{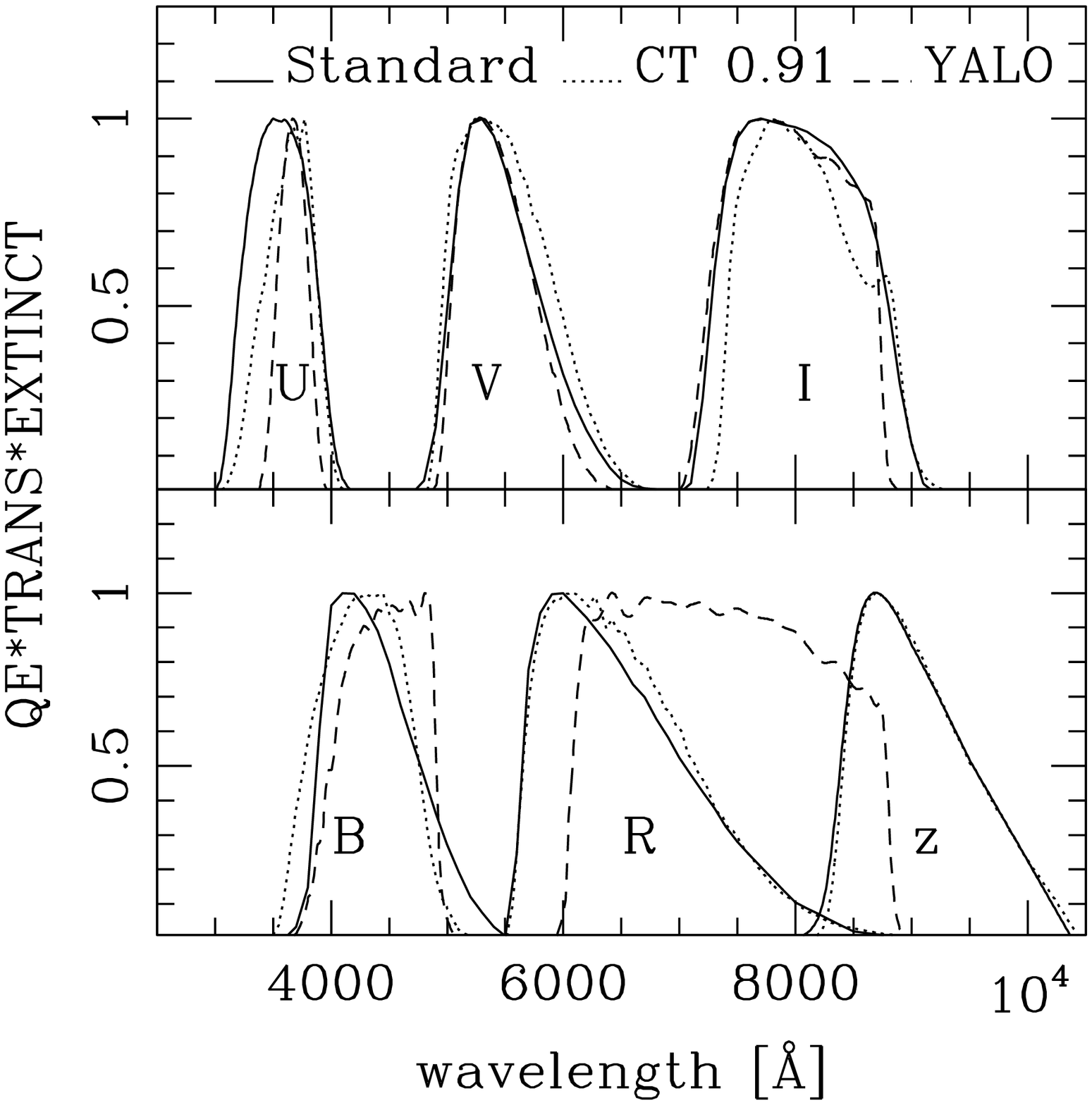}
\caption{Comparison of $UBVRIz$ bandpasses for the YALO (dashed line) and CTIO 0.91-m (dotted line) telescopes. Each of these
curves corresponds to the multiplication of the filter transmission curve, CCD quantum efficiency,
and the atmospheric continuum transmission spectrum for one airmass. Atmospheric line opacity
is not included in these bandpasses because they are used with spectra containing telluric
lines. Also shown (solid line) are the \citet{bessell90} standard Johnson/Kron-Cousins functions and the standard $z$ function defined
by \citet{hamuy01}.
\label{stritzinger.fig1}}
\end{figure}

Optical observations of IC~5179 began on 1999 October 8 and extended continuously 
through 1999 December 9, almost on a nightly basis. 
We collected the vast majority of data using the CTIO 0.91-m and YALO\footnote{YALO is a consortium
consisting of Yale Univ., Univ. of Lisbon, Ohio State Univ., and NOAO.} 1-m telescopes.
On the 0.91-m telescope we used the standard 3$\times$3 inch $UBV(RI)_{KC}$ Johnson/Kron-Cousins filter set
and a Gunn $z$ filter \citep{schneider83}, in combination with a Tek 2048$\times$2046 CCD detector which provided a scale
of 0.396 arcsec pix$^{-1}$. On the YALO telescope we used the dual-channel, optical-IR ANDICAM camera
which was equipped with a Loral 2048$\times$2048 CCD (0.3 arcsec pix$^{-1}$), a Stromgren $u$ filter 
and $BVRI$ filters. While the YALO $BVI$ filters were close to the standard system, the $u$ filter was narrower
than the Johnson bandpass and the $R$ filter was very wide compared to the standard bandpass.
Figure \ref{stritzinger.fig1} shows the filter bandpasses for both instruments compared
to the standard ones defined by \citet{bessell90} and \citet{hamuy01}.
We obtained a few additional photometric observations with the CTIO 1.5-m telescope
as well as with the ESO NTT and Danish 1.54-m telescopes.  Refer to Table \ref{journal} for the log of
the observations, which includes the date of observation, telescope used, observatory, and observer name(s).
	
\begin{figure}
\figurenum{2}
\epsscale{1.0}
\plotone{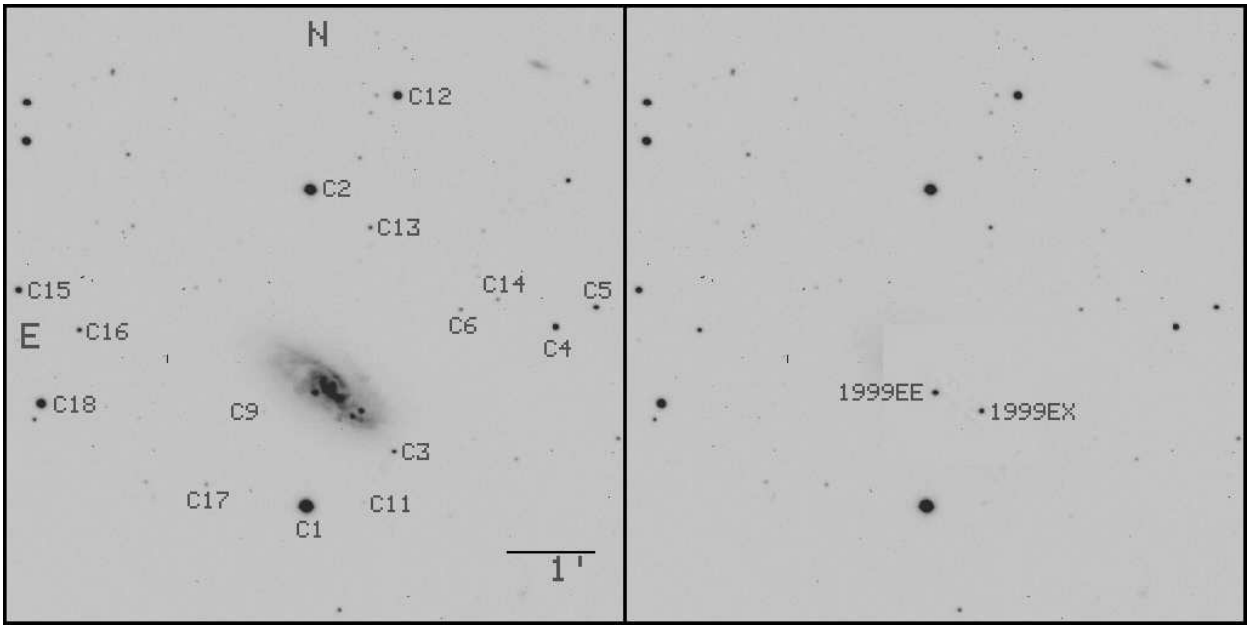}
\caption{$V$ band CCD image of SN~1999ee and SN~1999ex in IC~5179 taken on 1999 Nov. 17.
The photometric sequence stars are labeled. The right panel displays the galaxy subtracted
image clearly revealing both SNe. To avoid subtracting the local photometric sequence
from the subtracted image we restricted the subtraction to the image section around the host galaxy.
\label{stritzinger.fig2}}
\end{figure}

In order to reduce the effects due to atmospheric extinction we determined 
the brightness of each SN differentially with respect to a 
sequence of field stars in the region around IC~5179. 
Figure \ref{stritzinger.fig2} shows a $V$ band image identifying the photometric sequence.
We obtained absolute photometry of these comparison stars with the CTIO 0.91-m
telescope on 7 photometric nights in which we observed $UBVRI$ standards
from the list of \citet{landolt92} covering a wide range in brightness, color, and airmass.
On four nights we also observed $z$ band standards from the list of \cite{hamuy01}.
In order to transform instrumental magnitudes to the standard Johnson/Kron-Cousins
system we assumed linear transformations of the following form,

\begin{equation}
U = u - k_U X + CT_U (u - b) + ZP_U,
\label{ueqn}
\end{equation}
\begin{equation}
B = b - k_B X + CT_B (b - v) + ZP_B,
\label{beqn}
\end{equation}
\begin{equation}
V = v - k_V X + CT_V (b - v) + ZP_V,
\label{veqn}
\end{equation}
\begin{equation}
R = r - k_R X + CT_R (v - r) + ZP_R,
\label{reqn}
\end{equation}
\begin{equation}
I = i - k_I X + CT_I (v - i) + ZP_I,
\label{ieqn}
\end{equation}
\begin{equation}
Z = z - k_z X + CT_z (v - z) + ZP_z,
\label{zeqn}
\end{equation}

\noindent where $U,B,V,R,I,Z$ refer to magnitudes in the standard system,
$u,b,v,r,i,z$ refer to instrumental magnitudes measured
with an aperture of 14 arcsec in diameter (the same employed by Landolt),
$k_i$ to the atmospheric extinction coefficient, $X$ to airmass,
$CT_i$ to the color term, and $ZP_i$ to the zero-point of the transformation.
For each photometric night we used least-squares fits to solve for the photometric coefficients given above.
Except for the $U$ band the typical scatter in the transformation equations amounted
to 0.01-0.02 mag (standard deviation), thus confirming the photometric quality of the night. For $U$ the
spread was typically 0.03-0.04 mag.
Table \ref{sequence} contains the $UBVRIz$ photometry of the sequence stars identified 
in Figure \ref{stritzinger.fig2} along with the corresponding errors in the mean.

From the many different photometric nights of the SOIRS program we noticed that
the color terms of the CTIO 0.91-m telescope varied little with time, thus allowing us to adopt average values. This
offered the advantage of reducing the number of fitting parameters in equations \ref{ueqn}-\ref{zeqn}
during the SN reductions. For the CTIO 1.5-m, NTT, and Danish 1.54-m telescopes we were
also able to derive average color terms from observations of Landolt standards on
multiple nights. For the YALO, on the other hand, we did not have observations of
Landolt stars so we used the photometric sequence itself to solve for color terms.
From the many nights of data we could not see significant changes in these
coefficients, thus allowing us to adopt average values.
Table \ref{coefficients} summarizes the average values for each telescope and associated filters.
This table reveals that the color terms for the instrumental $U$ and $B$ bands are
generally large owing to the rapid sensitivity drop of CCDs toward blue wavelengths,
which effectively shifts the bandpasses to the red. Except for the YALO
$R$ filter the remaining color terms are close to zero, thus implying a good match of
the instrumental bands to the standard system. The large color term for the
YALO $R$ filter is clearly not unexpected owing to its non-standard width (see Figure \ref{stritzinger.fig1}).

Before performing photometry of the SNe, we subtracted late time templates of the parent galaxy
from all images in order to reduce light contamination using the method
described by \citet{filippenko86} and \citet{hamuy94a}. We obtained good-seeing moonless $UBVRIz$ images of IC~5179
with the CTIO 1.5-m telescope on 2001 July 16, nearly two years after
discovery of the SNe. We took three images of IC~5179 in $BVIz$ and two $RU$ images 
with an exposure time of 300 seconds. We combined these to create deeper master images of the
host galaxy for the subtraction process. The procedure used for galaxy subtraction consisted of four steps,
each of which was performed with IRAF\footnote{The Image Reduction and Analysis
Facility (IRAF) is maintained and distributed by the Association of Universities
for Research in Astronomy, under a cooperative agreement with the 
National Science Foundation.} scripts. The steps included: (1) aligning the two images
using field star positions (coordinate registration); (2) matching the point spread
function (PSF) of the two images (PSF registration); (3) matching the flux scale of
the two images (flux registration); (4) subtracting the galaxy image from the SN+galaxy
image. To avoid subtracting the local photometric sequence from the subtracted image
we restricted the subtraction to a small image section around the host galaxy. With this approach
the local standards remained unaffected by the above procedure and we could proceed
to carry out differential photometry for the SN and the local standards from the same image.

Upon completing galaxy subtraction we performed differential photometry for
all images of the SNe using either aperture photometry (with the IRAF task PHOT)
or PSF photometry (with the IRAF task DAOPHOT), depending on the SN brightness.
When photon noise exceeded 0.015 mag (the uncertainty in an individual photometric
measurement caused by non-Poissonian errors) for either SN, we used the more
laborious PSF technique. The weather in Chile was very good around 1999 October/November
and the seeing values at CTIO and ESO were always below 2.5 arcsec, and generally between 1-1.5 arcsec (FWHM).
Hence, we chose an aperture of 2 arcsec to calculate instrumental magnitudes
for the SNe and the comparison stars. Likewise, during PSF photometry we used a PSF fitting
radius of 2 arcsec. While this aperture kept the sky noise reasonably low, it included
a large fraction ($\sim$90\%) of the total stellar flux, thus minimizing
errors due to possible variations of the PSF across the image. For each frame we used the local
standards to solve for the photometric transformation coefficients (equations \ref{ueqn}-\ref{zeqn}).
The advantage of doing differential photometry is that all stars in a frame are observed with the same airmass
so that the magnitude extinction is nearly the same for all stars, thus reducing the number
of free parameters in the transformation equations\footnote{Note that we neglected here any color dependence
of the extinction coefficient. In the $B$ band where the second-order extinction is the highest
-- 0.02 mag per airmass per unit color -- we could have made an error of 0.01 mag 
because of this assumption.}.
Moreover, having fixed the color terms to the average values, the photometric transformation for
a single night only involved solving the zero point for each filter. Typically the scatter in the
photometric transformation proved $<$0.02 mag (standard deviation).

\section{Results}
\label{results.sec}

\subsection{SN~1999ee}

Table \ref{eephotometry} lists the resulting $UBVRIz$ photometry of SN~1999ee.
The quoted errors correspond to uncertainties owing to photon statistics. We adopted
a minimum error of 0.015 mag in an individual magnitude to account for uncertainties
other than the Poissonian errors. This value is the typical scatter that we
observe from multiple CCD observations of bright stars whose Poissonian errors are
negligibly small. This table specifies whether we obtained the magnitudes 
via PSF fitting (DAO) or aperture photometry (PHOT). We made
observations for a total of 50 nights.

\begin{figure}
\figurenum{3}
\epsscale{1.0}
\plotone{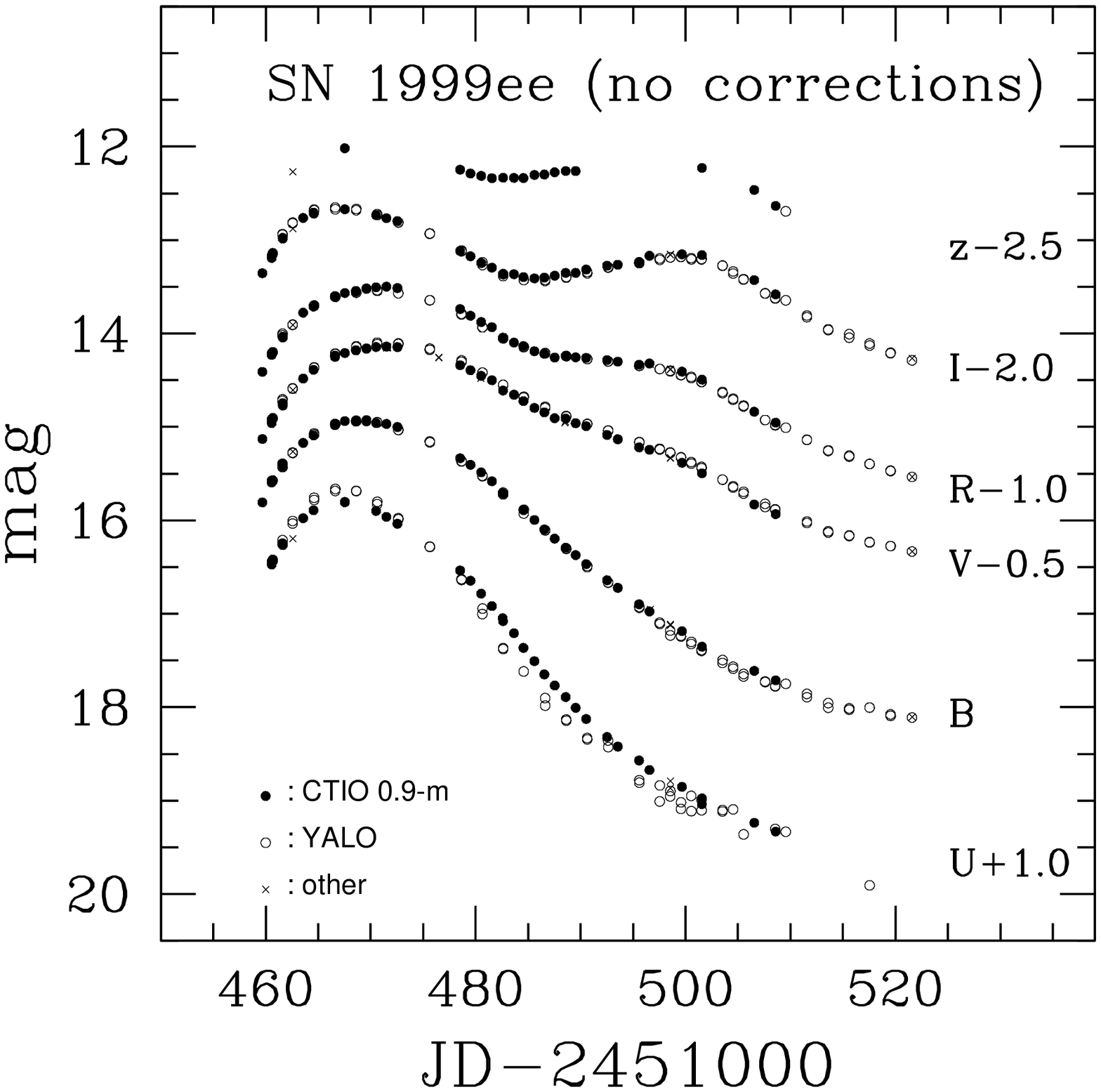}
\caption{$UBVRIz$ lightcurves of SN~1999ee. Open points and closed points show
photometry from the YALO and CTIO 0.91-m telescopes, respectively.
\label{stritzinger.fig3}}
\end{figure}

Figure \ref{stritzinger.fig3} presents the resulting lightcurves of SN~1999ee which reveal
the exceptional sampling obtained. This figure also shows
a systematic magnitude difference between the YALO and CTIO 0.91-m data 
in most bands.  This is a well known problem in SN observations caused by
the use of filters which do not exactly match each other
\citep{suntzeff88,menzies89,hamuy90,suntzeff00}.
Although the color terms derived for each filter have the specific purpose of
allowing us to standardize instrumental magnitudes, these are obtained
from stars with normal continuous spectra and are not expected to work
perfectly with SNe which are characterized by strong absorption and emission spectral features.
Figure \ref{stritzinger.fig3} shows that the systematic differences varied with time as the SN
evolved. They proved particularly pronounced in the $U$ band due to the
large departures of the instrumental filters from the Johnson bandpass.
Near maximum light, the difference in the photometry was
$\Delta(U,B,V,R,I)$=(-0.14,-0.01,-0.04,+0.04,-0.03) in the sense YALO minus CTIO 0.91-m.
One month later (JD 2,451,500) the difference was (+0.17,+0.04,-0.05,+0.03,+0.04).

We attempted to fix this problem and compute magnitude corrections 
by convolving the spectrophotometry of SN~1999ee \citep{hamuy02}
with instrumental and standard bandpasses.
Since we measured the SN magnitudes with photon detectors, a synthetic magnitude on the natural system 
must be calculated as the convolution of the object's photon number distribution ($N_\lambda$)
with the filter instrumental bandpass (S($\lambda$)), i.e.,

\begin{equation}
mag = -2.5~log_{10}~\int N_{\lambda}~S(\lambda)~d\lambda ~+~ZP,
\label{mageqn}
\end{equation}

\noindent where ZP is the zero-point for the magnitude scale.
To a minimum S($\lambda$) should include the transparency of the Earth's atmosphere, the filter transmission,
and the detector quantum efficiency (QE).  For the standard bandpasses
we adopted the $UBVRI$ filter functions given by \citet{bessell90} and
the $z$ function given by \citet{hamuy01}. However, since the Bessell curves are
meant for use with energy and not photon distributions
(see Appendix in \citet{bessell83}), it proved necessary to divide them by $\lambda$ before
employing them in equation \ref{mageqn} \citep{suntzeff99,hamuy01}.
For the instrumental bandpasses we used transmission
curves corresponding to the filters employed with the YALO and CTIO 0.91-m telescopes,
nominal QE curves for the LORAL and Tek CCDs used with both cameras,
and an atmospheric transmission spectrum corresponding to one airmass. Note that
we did not include here mirror aluminum reflectivities, the dichroic transmission
for the YALO camera, and dewar window transmissions because we lacked such information.

To check if the resulting bandpasses provided a good model for those actually
used at the telescopes we computed synthetic magnitudes for spectrophotometric
standards \citep{hamuy94b} and asked if the color terms derived
from the synthetic magnitudes matched the observed values given in Table \ref{coefficients}.
We could only do this calculation for the $BVRIz$ filters since the
spectrophotometric standards do not cover the entire $U$ filter. We calculated
synthetic color terms with equations \ref{beqn}-\ref{zeqn}, where $b,v,r,i,z$ are
the synthetic magnitudes for the YALO and CTIO 0.91-m instrumental bandpasses and
$B,V,R,I,Z$ are the synthetic magnitudes derived from the standard functions.
Although we find an overall good agreement between the observed and synthetic color terms,
small differences are present which suggest some differences between the nominal
and the instrumental bandpasses used at the telescopes.
Possible explanations for the differences are mirror reflectivities or any
transmissivity of other optical elements of the instruments employed
(e.g. the dichroic mirror used with YALO/ANDICAM). To solve this problem our approach consisted in applying
wavelength shifts to our nominal instrumental bandpasses until we were able to reproduce
exactly the measured color terms. The required shifts are summarized in Table \ref{shifts}.
They are always less than 100 \AA, except in the $V$ and $R$ YALO filters where they amount to 
120 \AA~and 180 \AA, respectively.  Figure \ref{stritzinger.fig1} displays the resulting functions.

Armed with the best possible models for our instrumental bands we 
used the spectrophotometry of SN~1999ee to compute magnitude 
corrections (S-corrections, hereafter) that should allow us to bring our 
observed photometry to the standard system.
For the $V$ filter the S-correction is  

\begin{equation}
\Delta V = V - v - CT_V(b-v) - ZP_V,
\label{magdiffeqn}
\end{equation}

\noindent where $V$ is the SN synthetic magnitude computed with the Bessell function, and
$b$,$v$ are the SN synthetic magnitudes computed with the instrumental bandpasses.
$CT_V$ is the color term for the corresponding filter and $ZP_V$ is a zero point
that can be obtained from the spectrophotometric standards to better than 0.01 mag. 
Similar equations can be written for the other filters.

\begin{figure}
\figurenum{4}
\epsscale{1.0}
\plotone{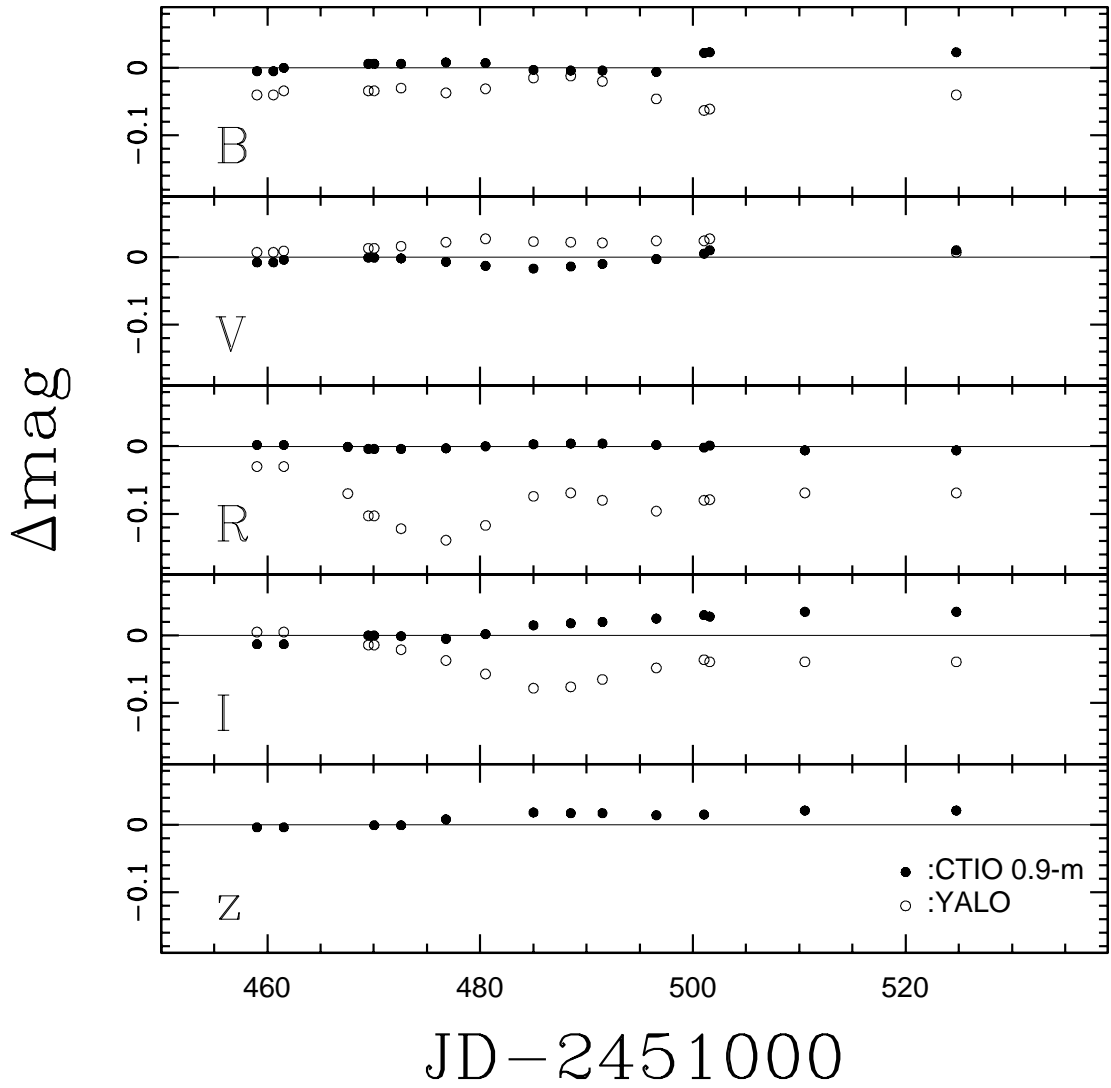}
\caption{S-corrections as a function of Julian Date for SN~1999ee, computed from the YALO (open  points) and CTIO 0.91-m (closed points).
\label{stritzinger.fig4}}
\end{figure}

Figure \ref{stritzinger.fig4} shows the S-corrections.                 
For the CTIO 0.91-m system
the corrections are generally small, which suggests that this instrument provides
a good match to the standard system. The worst case occurs in the $I$ band at late times where
the shifts amount to 0.04 mag. The YALO instrument, on the other hand,
is affected by large systematic shifts of 0.05 mag or more. The S-corrections
are particularly large in the $R$ filter, which is not unexpected considering
the large departures from the standard band.

\begin{figure}
\figurenum{5}
\epsscale{1.0}
\plotone{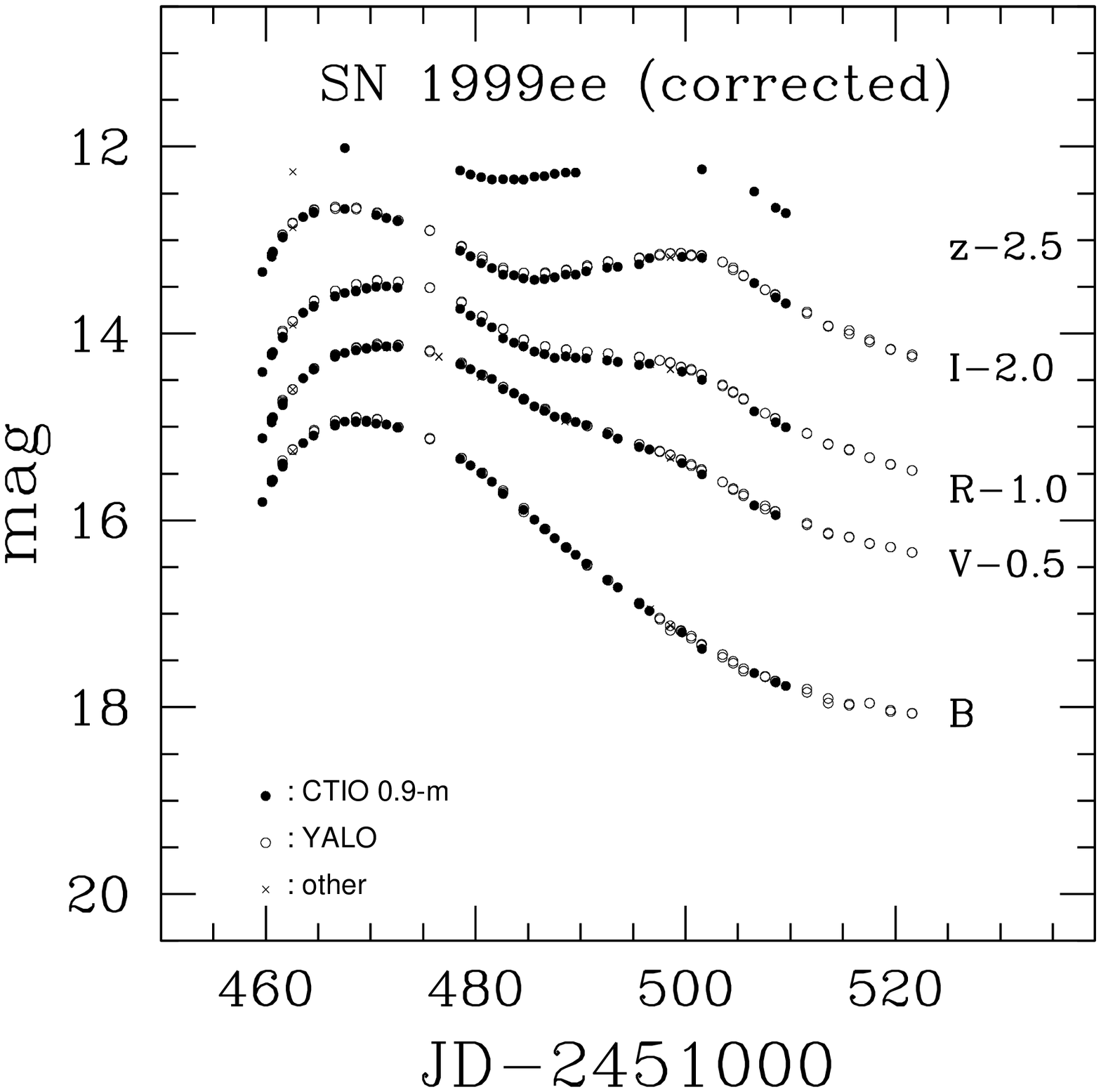}
\caption{$UBVRIz$ lightcurves of SN~1999ee after applying S-corrections.
Open points and closed points show photometry from the YALO and CTIO 0.91-m telescopes, respectively.
\label{stritzinger.fig5}}
\end{figure}

Figure \ref{stritzinger.fig5} contains the $BVRIz$ lightcurves after applying
S-corrections. A comparison with Figure \ref{stritzinger.fig3}
shows that the corrections help significantly in the $V$ band at all epochs.
In the  $B$ and $I$ bands they only help at some epochs, while 
it is clear that we overcorrect in the $R$ band at all epochs. These differences
are somewhat disappointing and are certainly due to an incomplete knowledge of the
actual bandpasses employed at the telescopes. We explored if the problem could be
due to a poor flux calibration of the SN spectra or to incorrect reddening corrections.
For this purpose we applied a $\pm$0.1 $E(B-V)$ artificial reddening to the SN spectra and
re-computed the S-corrections. The result of this test is that the S-corrections
change only by 0.005 mag in each direction, which implies that they
are mostly determined by spectral features that vary on small wavelength scales
compared to the bandpasses.

Although we do not feel we should
use these S-corrections at this stage, this exercise demonstrates the need of
accurately characterizing the instrumental system before great photometric accuracies
of SNe~Ia can be claimed. It also shows that this effect can have important astrophysical consequences
since a small systematic shift of 0.02 mag in the $B-V$ color can introduce
a $0.08$ mag error in the extinction corrected peak $B$ magnitudes of a SN and thus
lead to biased cosmological parameters.

Returning to the uncorrected magnitudes of SN~1999ee, we used a high order Legendre polynomial
to derive $B_{max}$ of 14.93$\pm$0.02 mag on JD 2,451,469.1$\pm$0.5, which indicated that the first
observations began 10 days prior and the last observations were made 53 days after $B_{max}$.
Likewise, we found $V_{max}$ of 14.61$\pm$0.03 mag about 2 days after $B_{max}$, in agreement
with the template lightcurves derived by \cite{leibundgut91}.
All filters covered the evolution of SN~1999ee extremely well through maximum light,
past the secondary maximum through the onset of the exponential tail due to
radioactive decay of $^{56}Co$$\rightarrow$$^{56}Fe$. The $z$ band lightcurve 
is the first of its type for a SN~Ia event and reveals that the secondary maximum
was only a few tenths of a magnitude fainter that the primary peak. Lightcurve parameters
for all filters are given in Table \ref{lcpar}. We estimated the uncertainties 
in the peak magnitudes from the scatter (peak to peak) in the photometric points
around maximum in order to account for S-corrections.

\begin{figure}
\figurenum{6}
\epsscale{1.0}
\plotone{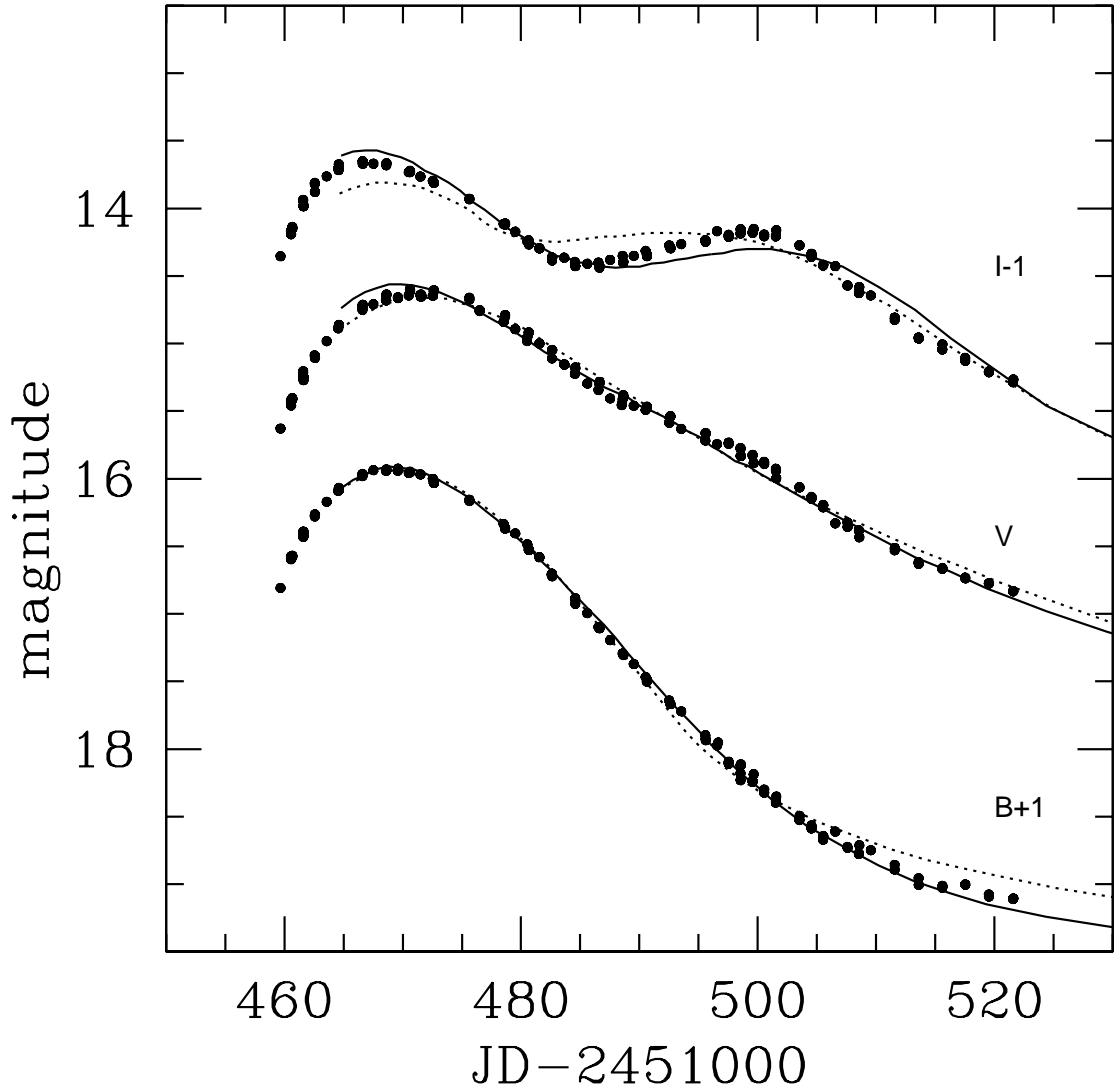}
\caption{
Template fitting for SN~1999ee. Solid line is SN~1992bc and
dashed line is SN~1991T, both modified for time dilation and K correction for the corresponding 
redshift of SN~1999ee.
\label{stritzinger.fig6}}
\end{figure}

We next performed lightcurve fits for this event using the templates derived by \citet{hamuy96}.
Figure \ref{stritzinger.fig6} shows the $BVI$ fits obtained with the 1991T and 1992bc templates
that best matched the observed lightcurves. We stretched both templates by (1+$z$) in order
to account for time dilation, and modified them by K terms appropriate for the redshift of SN~1999ee
\citep{hamuy93}.  While the fits are reasonably good in the $B$ and $V$ bands, the templates
do not provide a good match to the observed $I$ lightcurve. The small decline rates
of these templates ($\Delta$$m_{15}(B)$=0.87 for SN~1992bc and 0.94 for SN~1991T) reveal that
SN~1999ee was a slow-declining SN. In fact, a direct measurement of the decline
rate gave $\Delta$$m_{15}(B)$=0.91$\pm$0.04 which, after correction for host-galaxy extinction
\citep{phillips99} increased to 0.94$\pm$0.06 mag (see section \ref{99ee.sec}).

\subsection{SN~1999ex}

\begin{figure}
\figurenum{7}
\epsscale{1.0}
\plotone{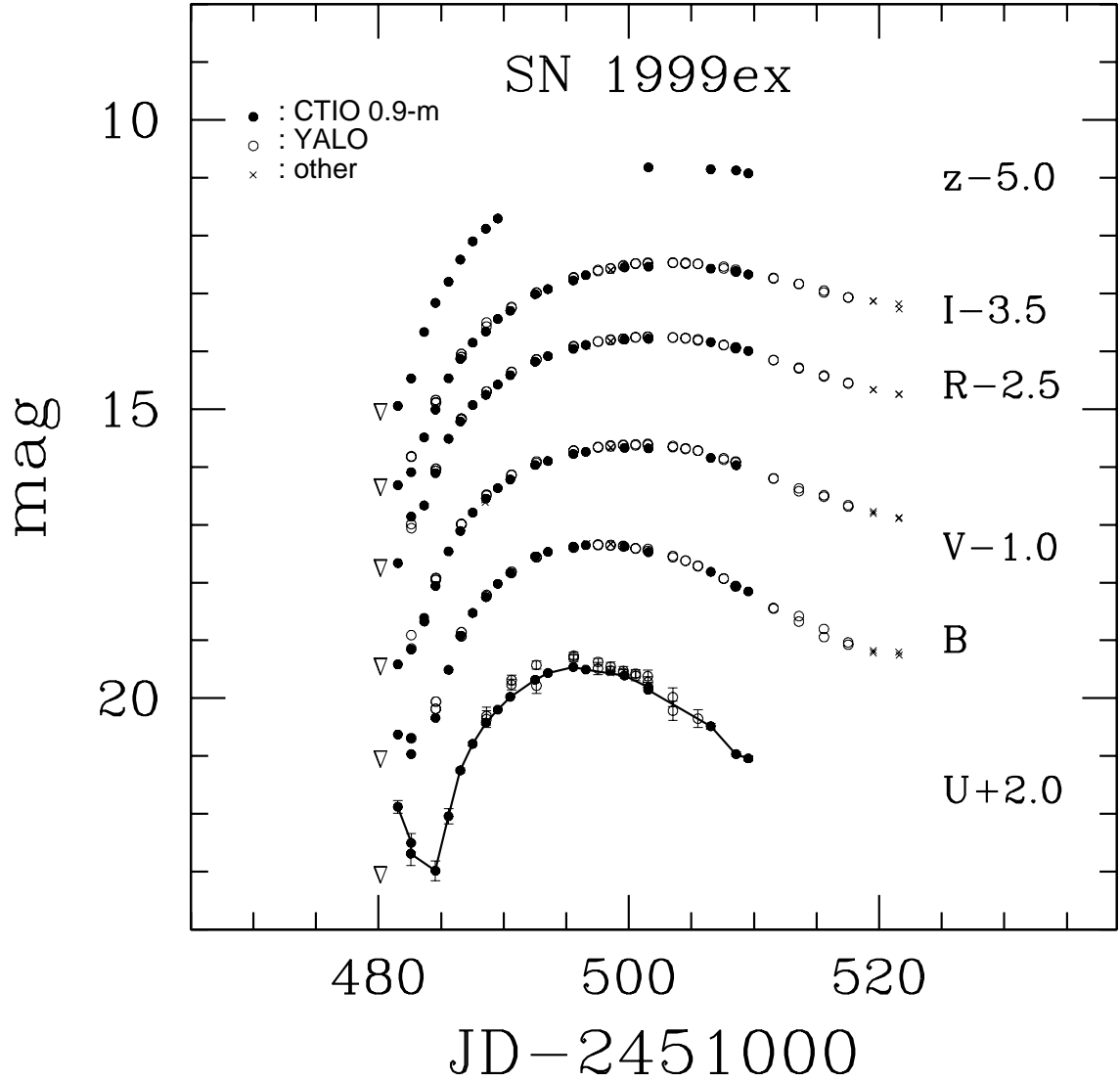}
\caption{$UBVRIz$ lightcurves of SN~1999ex. Open points and closed points show
photometry from the YALO and CTIO 0.91-m telescopes, respectively. Upper limits 
derived from images taken on JD 2,451,480.5 are also included. The solid line through
the $U$ data is drawn to help the eye to see the initial upturn.
\label{stritzinger.fig7}}
\end{figure}

Table \ref{exphotometry} contains the $UBVRIz$ photometry for SN~1999ex, as well as the method used
to determine magnitudes. We present here a total of 32 nights of data.
The lightcurves are presented in Figure \ref{stritzinger.fig7}. Clearly the observations
began well before maximum light thanks to our continuous followup of the host galaxy owing to
the prior discovery of the Type Ia SN~1999ee. The first detection occurred on JD 2,451,481.6 (1999 October 30) in all filters.
Excellent seeing images obtained on the previous night allowed us to place reliable upper limits to the SN
brightness. Hence, in what follows we assume that shock breakout took place on JD 2,451,480.5.
Given our nightly observations of IC~5179 we can attach an uncertainty of $\pm$0.5 day to this estimate.
Undoubtedly, these are the earliest observations of a SN~Ib/c, which reveal the
SN evolution right after explosion. The most remarkable feature in this figure is the early dip in the
$U$ and $B$ lightcurves -- covering the first 4 days of evolution -- after which the SN steadily rose
to maximum light. The dip is absent in $VRIz$, yet it is possible to notice a slight change in the rate
of brightening at the earliest epochs. This early upturn is reminiscent of the Type II SN~1987A \citep{hamuy88} 
and the Type IIb SN~1993J \citep{schmidt93,richmond94}, both of which showed a first phase of rapid dimming
followed by a second phase of steady brightening.  Until now this dip had never been observed in a SN~Ib/c.
Although this is a generic feature of core collapse or thermonuclear explosion models, in
Section \ref{99ex.sec} we argue that these observations favor the core collapse nature of SN~Ib/c.

SN~1999ex reached $B_{max}$=17.35$\pm$0.02 on JD 2,451,498.1 and $V_{max}$=16.63$\pm$0.04 
3 days later on JD 2,451,501.2. This implies that our observations began 17 nights before $B_{max}$
and extended until 24 days after peak. The upper limits on JD 2,451,480.5 strongly suggest that 
shock breakout occurred 18 days before $B_{max}$. Table \ref{lcpar} summarizes peak magnitudes
and time of occurrence for each filter. 

\begin{figure}
\figurenum{8}
\epsscale{1.0}
\plotone{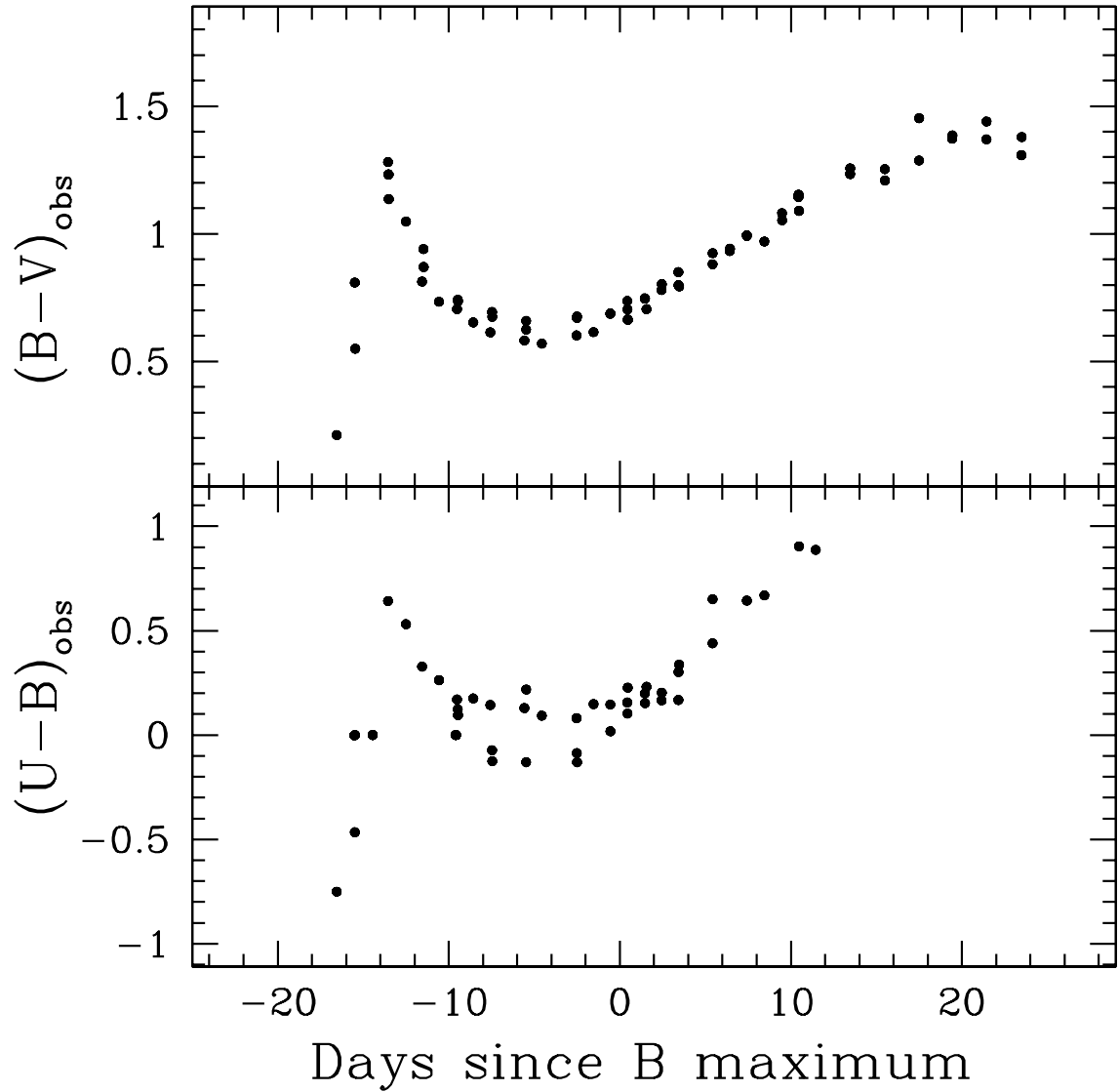}
\caption{$B-V$ and $U-B$ color curves of SN~1999ex as a function of time since $B$ maximum.
\label{stritzinger.fig8}}
\end{figure}

Figure \ref{stritzinger.fig8} shows the $U-B$ and $B-V$ color curves for SN~1999ex, both of which
track photospheric temperature variations.
Initially the photosphere displayed a rapid cooling in which the $B-V$ color increased from 0.2 to 1.3
in only four days. Then the photospheric temperature increased so the SN evolved back to the blue. 
This phase extended for ten days through maximum light when SN~1999ex reached $B-V$=0.6.
After reaching maximum light the SN again turned toward the red and the photosphere began to cool.
The observations ceased 24 days past $B_{max}$. At that time there seemed to be an
inflection of the $B-V$ curve toward the blue which probably coincided with the onset
of the nebular phase when the SN went from being optically thick to optically thin.

\section{Discussion}
\label{disc.sec}
\subsection{SN~1999ee}
\label{99ee.sec}

In order to obtain absolute magnitudes for SN~1999ee we first need to correct the
apparent magnitudes for the effects of reddening caused by dust in both our Galaxy
and the host galaxy. A small reddening of $E(B-V)_{Gal}$=0.02 can be attributed to our Galaxy in the
direction of IC~5179 according to the IR dust maps of \citet{schlegel98}.
Interstellar absorption lines due to Na I D $\lambda$$\lambda$5890,5896, Ca II $\lambda$3934,
and Ca II $\lambda$3968 at the redshift of IC~5179 could be clearly seen in the spectra
of SN~1999ee, with equivalent widths (EW) of 2.3, 0.8, and 0.6 \AA, respectively \citep{hamuy02}. This
suggests that a non-negligible amount of absorption in IC~5179 affected SN~1999ee.
Calibrations between EW and reddening have been derived from high resolution spectra
of Galactic stars \citep{richmond94,munari97}, but these are only valid in the
non-saturated regime (EW(Na I D)$<$0.8 \AA). Using SN data \cite{barbon90} derived the relation
$E(B-V)$ $\sim$ 0.25 EW (Na I D), which implies $E(B-V)_{host}$=0.58 for SN~1999ee.
We believe, however, that this value is of little usefulness owing 
to the uncertain dust to gas ratio around SNe \citep{hamuy01}.

Instead we can estimate the host galaxy's reddening of SN~1999ee 
using the method described by \citet{phillips99} and \cite{lira95}.
The de-reddening technique consists in comparing the observed colors of the SN
to the intrinsic ones which are a function
of the decline rate of the SN. $\Delta$$m_{15}(B)$ itself can be affected by the
extinction due to the change of the effective wavelength of the $B$ filter
as the SN spectrum evolves \citep{phillips99}. From the direct measurement of $\Delta$$m_{15}(B)_{obs}$=0.91$\pm$0.04
(after correction for time dilation and K-terms) and an internal 
reddening of $E(B-V)_{host}$=0.28$\pm$0.04 (see below),
we derive a true decline rate of $\Delta$$m_{15}(B)$=0.94$\pm$0.06. 

A direct comparison between the intrinsic colors $(B-V)_0$ and $(V-I)_0$ as 
predicted by \citet{phillips99} and the observed colors at maximum light 
(after correction for Galactic reddening, time dilation,
and K-terms) yields $E(B-V)_{max}$=0.39$\pm$0.11 and $E(V-I)_{max}$=0.29$\pm$0.11.
In addition using colors from the exponential tail we determine $E(B-V)_{tail}$=0.27$\pm$0.05.
Taking the weighted mean of $E(B-V)_{max}$, $E(B-V)_{tail}$, and 0.8$E(V-I)_{max}$ we obtain  
the final estimation for host galaxy reddening $E(B-V)_{host}$=0.28$\pm$0.04.
Including Galactic extinction we get $E(B-V)_{total}$=0.30$\pm$0.04 which, following 
\citet{phillips99}, leads to the following monochromatic absorption values:
$A_{B}$=1.24$\pm$0.17, $A_{V}$=0.94$\pm$0.13, and $A_{I}$=0.55$\pm$0.07.

After applying these corrections to our $BVI$ peak apparent magnitudes 
and using the distance modulus of 33.55$\pm$0.23
based on the galaxy redshift, we obtain final absolute peak magnitudes of
$M_{B}$=-19.85$\pm$0.28, $M_{V}$=-19.87$\pm$0.26, and $M_{I}$=-19.43$\pm$0.24.
For its decline rate of $\Delta$$m_{15}(B)$=0.94$\pm$0.06, SN~1999ee fits
well with the peak luminosity-decline rate relation derived by \citet{phillips99}.

Instead of adopting a distance to solve for absolute magnitudes, we can
now solve for the distance to IC~5179 using the absolute magnitudes
of SNe~Ia recently calibrated with Cepheids. Note that we are not using the
previous redshift based absolute magnitudes so that this is an independent test.
The Cepheid-based absolute magnitudes (corrected for reddening and normalized to
a $\Delta$$m_{15}(B)$=1.10) of the SNe~Ia calibrators used in \citet{phillips99}
are $M_{B}$=-19.61$\pm$0.09, $M_{V}$=-19.58$\pm$0.08, and $M_{I}$=-19.29$\pm$0.10.
For $\Delta$$m_{15}(B)$=0.94$\pm$0.06, the luminosity-decline
rate relation mentioned above predicts absolute magnitudes of $M_{B}$=-19.72$\pm$0.12,
$M_{V}$=-19.67$\pm$0.11, and $M_{I}$=-19.34$\pm$0.12 for SN~1999ee. From the
peak apparent magnitudes we
obtain distance moduli of $\mu_{B}$=33.42$\pm$0.21, $\mu_{V}$=33.35$\pm$0.17, 
and $\mu_{I}$=33.46$\pm$0.14 hence yielding $\mu_{avg}$=33.42$\pm$0.10 for IC~5179.
This value compares well with $\mu$=33.55$\pm$0.23 derived from the galaxy redshift.

\begin{figure}
\figurenum{9}
\epsscale{1.0}
\plotone{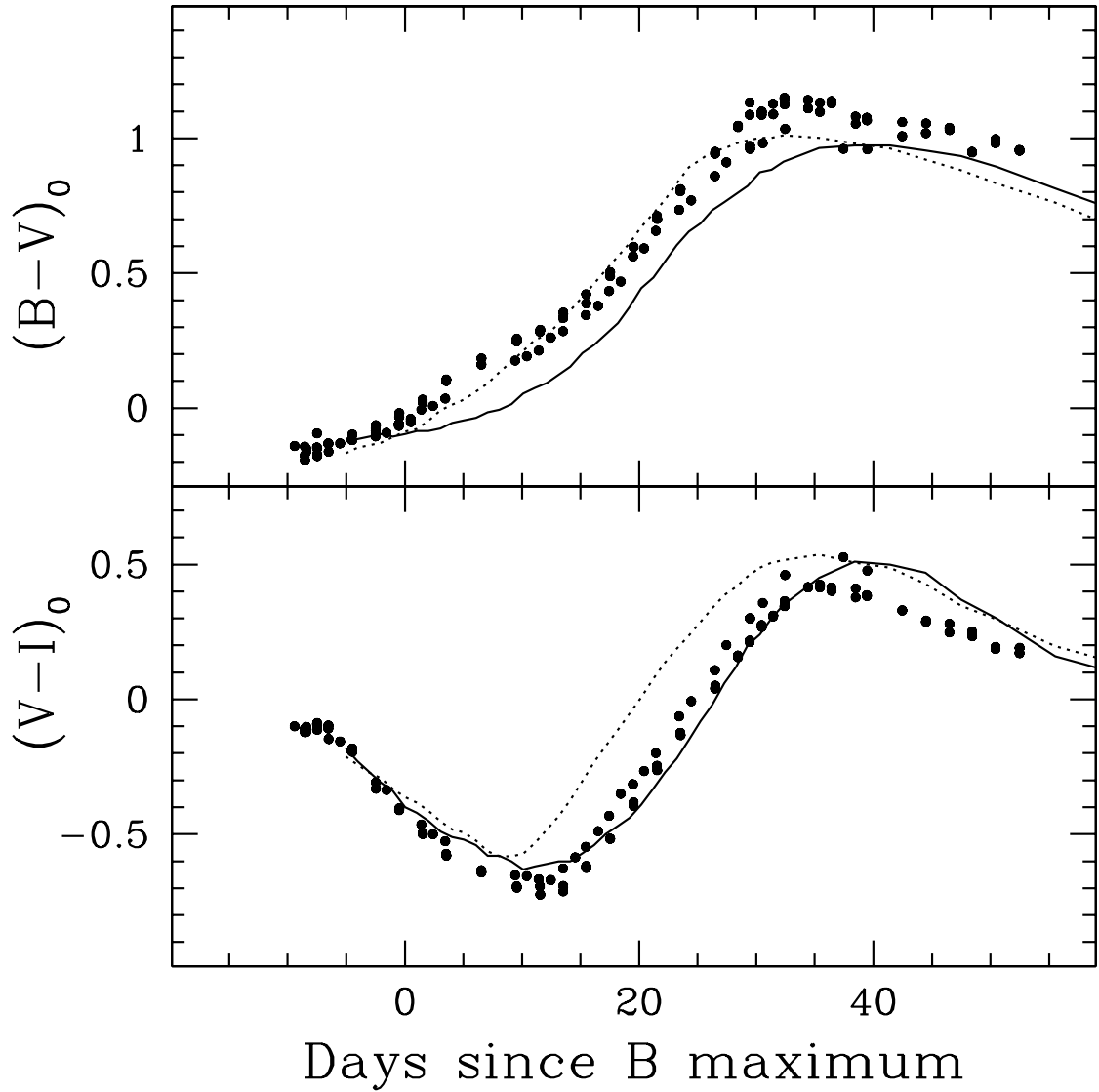}
\caption{$(B-V)_0$ and $(V-I)_0$ color curves of SN~1999ee as a function of time since $B$ maximum.
For comparison are shown the templates curves of SN~1992bc (solid line) and SN~1991T (dotted line).
\label{stritzinger.fig9}}
\end{figure}

Figure \ref{stritzinger.fig9} shows the $(B-V)_0$ and $(V-I)_0$ color curves of
SN~1999ee corrected for $E(B-V)_{total}$=0.30, compared to the template curves of the slow-declining
SNe~1992bc (solid line) and SN~1991T (dotted line), both adjusted to match the colors of SN~1999ee at maximum
light, time dilated, and K-corrected. This comparison reveals that, although SN~1999ee had the typical
trends displayed by other SNe~Ia, it is not possible to fit its color at all epochs with a single reddening correction.
SN~1992bc seems to provide a better match in $V-I$ before day +30 but after that epoch there
is clear mismatch. SN~1991T, on the other hand, provides a good fit in $B-V$ yet a
poor match in $V-I$. This demonstrates that SNe~Ia with nearly the same
decline rate do not have identical color curves. These examples raise doubts about
the derivation of reddening from the shapes of the color curves, as advocated by
\cite{riess96}. The method implemented by \cite{phillips99}, instead, does not use information on
the shape of the color curve but only the color at individual epochs.

\subsection{SN~1999ex}
\label{99ex.sec}

SN~1999ex also showed clear absorption features due to interstellar lines of 
Na I D $\lambda$$\lambda$5890,5896, Ca II $\lambda$3934, and Ca II $\lambda$3968 at the redshift of IC~5179.
The equivalent widths of 2.8, 1.8, and 0.9 \AA~measured for these lines \citep{hamuy02} suggest
even more dust absorption than for SN~1999ee.
As mentioned above, however, it proves difficult to derive dust absorption from interstellar lines
owing to the very uncertain dust to gas ratio around SNe. A lower limit to the
peak absolute magnitudes of SN~1999ex can be derived assuming $E(B-V)$$_{Gal}$=0.02 \citep{schlegel98}.
Omitting K-corrections (which are expected to be $<$0.02 mag, assuming the K-terms for SNe~Ib/c
are similar for SNe~Ia), we get $M_{B}$$<$-16.28$\pm$0.23, $M_{V}$$<$-16.98$\pm$0.23, and $M_{I}$$<$-17.61$\pm$0.23 for SN~1999ex.
Assuming that the amount of reddening affecting
SN~1999ex is not radically different than $E(B-V)$$_{host}$=0.28$\pm$0.04
derived from SN~1999ee we get absolute magnitudes which may be considered
representative of what the actual reddening might give.
With this assumption we obtain $M_{B}$$=$-17.44$\pm$0.28, 
$M_{V}$$=$-17.86$\pm$0.26, and $M_{I}$$=$-18.12$\pm$0.24. 
These calculations illustrate how including
host galaxy reddening affect the absolute magnitudes
and, perhaps more importantly, the bolometric lightcurve (see below).

Although the initial dip and UV excess observed in SN~1999ex had never been observed
before among Ib/c events, it was also observed in the Type II SN~1987A \citep{hamuy88} and
the Type IIb SN~1993J \citep{schmidt93,richmond94}. For these SNe~II it is thought that the initial dip
corresponded to a phase of adiabatic cooling that ensued the initial UV flash caused by
shock emergence which super-heated and accelerated the photosphere. The following
brightening is attributed to the energy deposited behind the photosphere by the radioactive
decay of $^{56}Ni$$\rightarrow$$^{56}Co$ and $^{56}Co$$\rightarrow$$^{56}Fe$.
Although the lightcurve of SN~1999ex bears qualitative 
resemblance to SN~1987A and SN~1993J, in those cases the initial peaks were relatively brighter
and the dips in the lightcurves occurred $\sim$8 days after explosion compared to 4 days in SN~1999ex.
This suggests that SN~1999ex also was a core collapse SN and that the photometric
differences could be due to the hydrogen that the progenitors of SN~1987A and SN~1993J were
able to retain.  \citet{woosley87} computed Type Ib SN models consisting of the explosion
of a 6.2 M$_\odot$ helium core. Their Figure 7 shows the bolometric
luminosities of three models with different explosion energies and
$^{56}$Ni nucleosynthesis. Despite the different lightcurve shapes and
peak luminosities, all three models show an initial peak followed by
a dip a few days later and the subsequent brightening caused by
$^{56}$Ni $\rightarrow$ $^{56}$Co $\rightarrow$ $^{56}$Fe, making them
good models for SN~1999ex.

\begin{figure}
\figurenum{10}
\epsscale{1.0}
\plotone{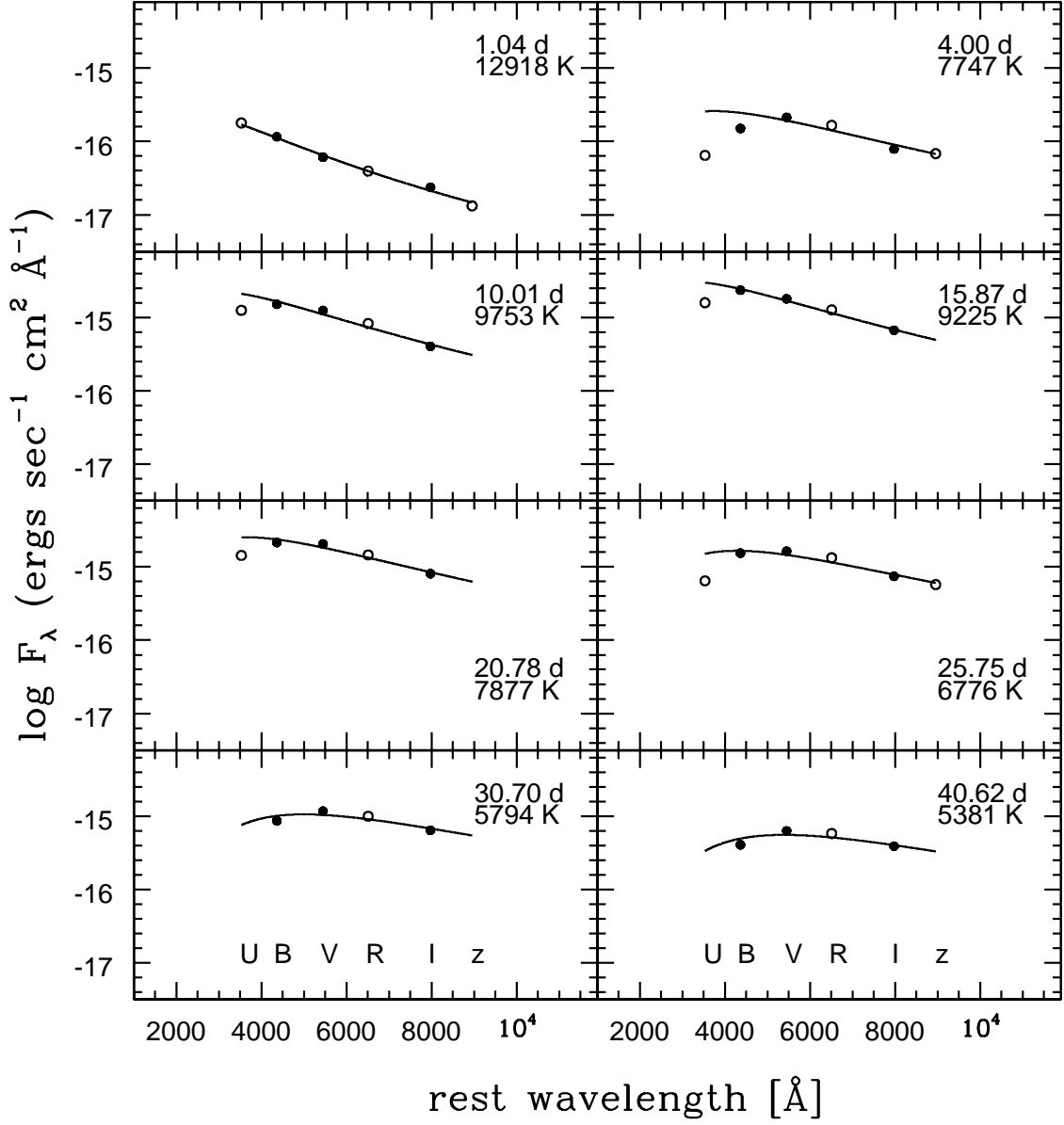}
\caption{Blackbody fits to the $BVI$ magnitudes (closed circles) of SN~1999ex for eight
representative epochs (measured since explosion time). The $URz$ magnitudes are also shown
(open circles) but they were not included in the fits. Although the fits were done by
converting BB fluxes into broad-band Vega magnitudes, for plotting purposes the SN 
magnitudes were converted into monochromatic fluxes (assuming $E(B-V)$$_{Gal}$=0.02 and $E(B-V)$$_{host}$=0.28).
\label{stritzinger.fig10}}
\end{figure}

\begin{figure}
\figurenum{11}
\epsscale{1.0}
\plotone{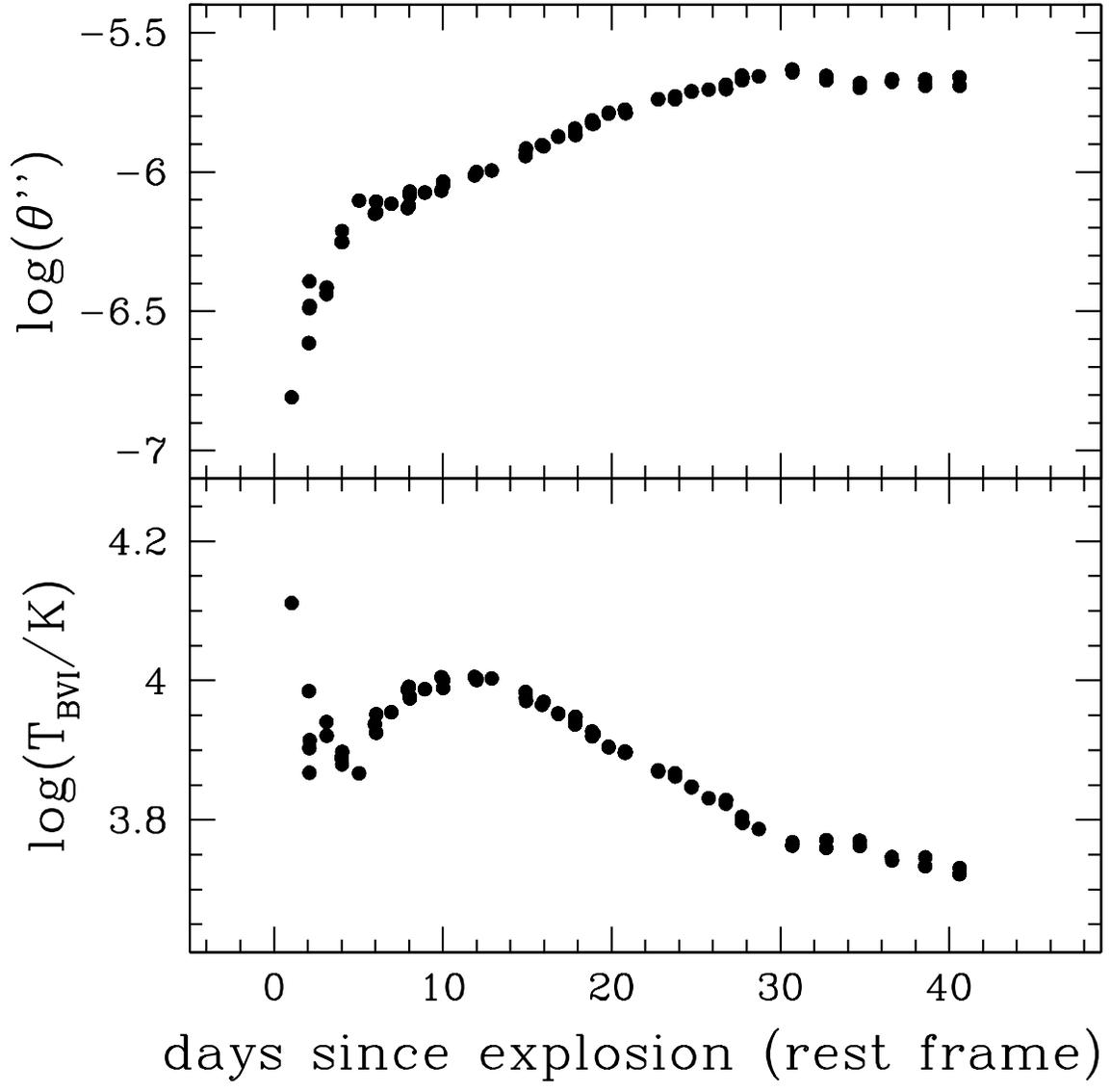}
\caption{Angular radius and color temperature evolution of SN~1999ex
derived from blackbody fits to the $BVI$ magnitudes, assuming $E(B-V)$$_{Gal}$=0.02,
$E(B-V)$$_{host}$=0.28, and JD 2,451,480.5 for the explosion time.
\label{stritzinger.fig11}}
\end{figure}

\begin{figure}
\figurenum{12}
\epsscale{1.0}
\plotone{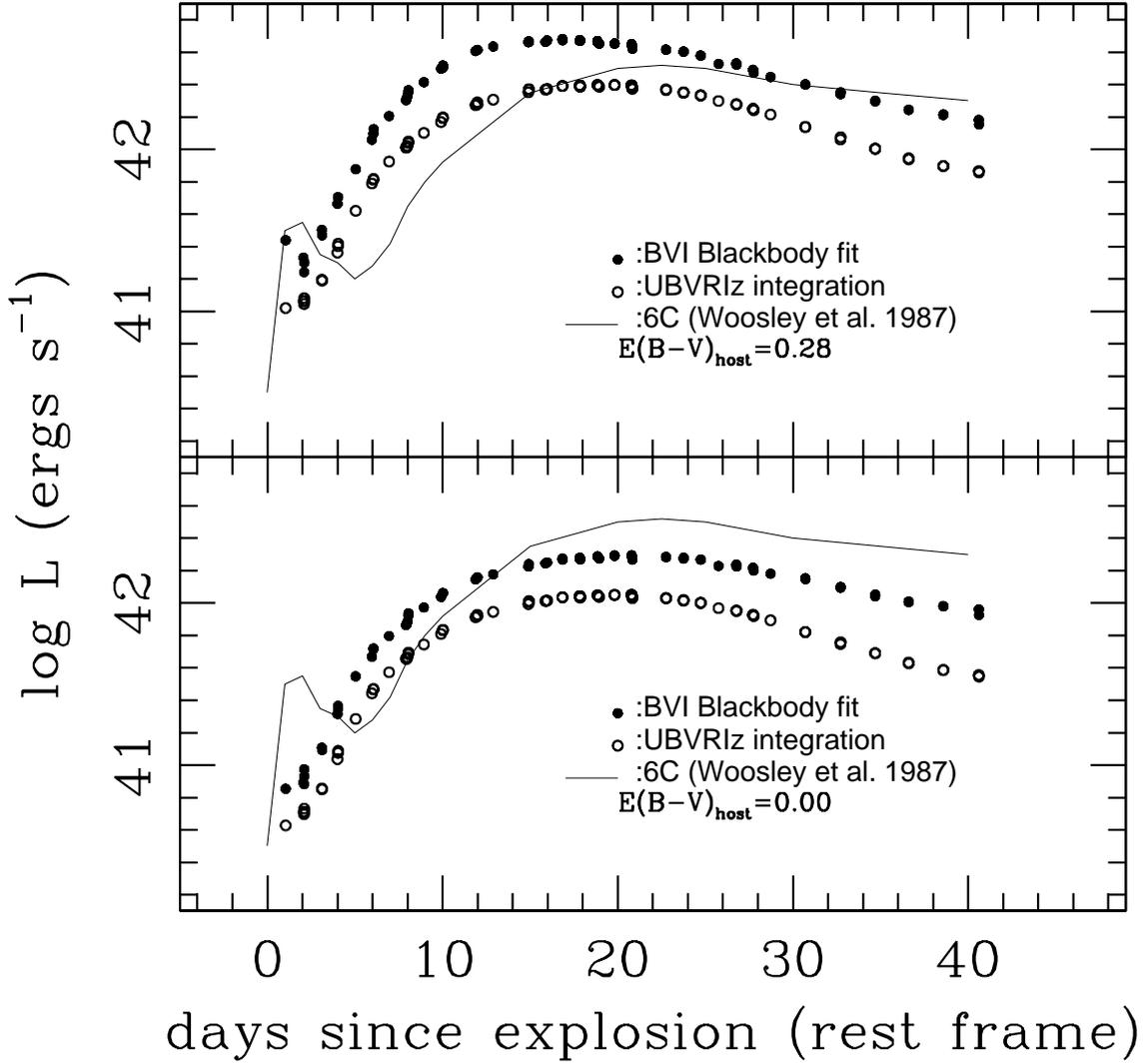}
\caption{Bolometric lightcurve of SN~1999ex derived from blackbody fits to the $BVI$
magnitudes (closed circles) and from a direct integration of the $UBVRIz$ fluxes (open points).
The top panel assumes $E(B-V)$$_{host}$=0.28 whereas the bottom panel adopts $E(B-V)$$_{host}$=0.00.
The solid line is the 6C hydrogenless core bounce SN model of \citet{woosley87}.
\label{stritzinger.fig12}}
\end{figure}

We proceed now to compute a bolometric lightcurve for SN~1999ex in order
to carry out a comparison with the models. A first approach consists in
performing a blackbody (BB) fit at each epoch and derive analytically the
total flux from the corresponding Planck function. Our fitting procedure consists
in computing synthetic magnitudes in the Vega system \citep{hamuy01} for BB
curves reddened by $E(B-V)$$_{Gal}$=0.02 and $E(B-V)$$_{host}$=0.28, performing
a least-squares fit to the SN $BVI$ magnitudes, and deriving the color
temperature and angular radius that yield the best match to the observed photometry.
Figure \ref{stritzinger.fig10} illustrates the fits for eight representative epochs.
Note that, while the BB curve provides excellent fits during the first days of SN evolution,
by day 4 since explosion the observed $U$ flux begins to fall below the BB curve.
This is a result of the strong line blanketing at these wavelengths,
assuming the opacities for SNe~Ib/c are similar to SNe~Ia \citep{pinto00}.
Note also that, although the $R$ and $z$ magnitudes are not included in the fits, they
match well the BB curves derived from the $BVI$ photometry. Table \ref{tal} summarizes
the parameters yielded by the BB fits for all epochs. Figure \ref{stritzinger.fig11} 
shows that the  photospheric angular radius showed a steady increase which reflects the
expansion of the SN ejecta. During the first 4 days the photospheric temperature dropped
due to the adiabatic cooling that followed shock breakout, after which the temperature
increased due to radioactive heating and then dropped again owing to expansion. Note that,
as expected, this curve bears close resemblance to the color curves shown in Figure \ref{stritzinger.fig8}.
The resulting bolometric lightcurve is plotted in the top panel of Figure \ref{stritzinger.fig12}
(closed circles) and the corresponding values can be found in Table \ref{tal}. 
For comparison the bottom panel shows the curve obtained by assuming $E(B-V)$$_{host}$=0.00.

An alternative route to derive bolometric fluxes is to perform a direct integration of the
SN $UBVRIz$ broad-band magnitudes. For this purpose we use the $UBVRIz$ magnitudes of Vega
\citep{hamuy92,hamuy01} and its spectral energy distribution \citep{hayes85} to derive conversion
factors between broad-band magnitudes and monochromatic fluxes for the standard 
photometric system shown in Figure \ref{stritzinger.fig1}. With these assumptions a zero magnitude
star has monochromatic fluxes of (3.98,6.43,3.67,2.23,1.17,0.82)$\times$10$^{-9}$
ergs sec$^{-1}$ cm$^{-2}$ \AA$^{-1}$ for $U,B,V,R,I,z$, which have equivalent wavelengths of 
3570, 4413, 5512, 6585, 8068, and 9058 \AA, respectively. The resulting $U-z$
bolometric fluxes are listed in Table \ref{tal} and shown with open circles in
Figure \ref{stritzinger.fig12}, both for $E(B-V)$$_{host}$=0.28 (top) and $E(B-V)$$_{host}$=0.00 (bottom).
Since the SN emitted more flux beyond the $U-z$ wavelength range, this approach is expected
to give a lower limit to the actual luminosity. Not surprisingly, the resulting fluxes lie
well below the BB fluxes. The BB fits to the $BVI$ magnitudes may also underestimate 
the true flux by a few dex owing to line blanketing in the $B$ band. 
On the other hand, flux redistribution of the missing $B$ flux could 
show up in the $VI$ bands and compensate for this flux deficit. 
Without a detailed atmosphere model for SNe~Ib/c it proves difficult
to quantify the difference between the BB and the true SN luminosity.
In the case of SN~1987A -- the SN with the best wavelength
coverage -- the $BVI$ BB fit yields a flux $\sim$0.05 dex greater than
the $U$$\rightarrow$$M$ bolometric flux of \cite{suntzeff90}, throughout
the optically thick phase of the SN. We believe therefore that
the BB fits to SN~1999ex are probably within 0.1 dex from the true 
bolometric luminosity.

Figure \ref{stritzinger.fig12} (top) reveals that the major difference between the 
BB and the $U-z$ curves is that the initial dip is not present
in the $U-z$ integration. This is due to the fact that the SN photosphere
was initially hot so that a significant fraction of the total flux
was emitted at wavelengths shorter than 3500 \AA. The BB fit on the other
hand, provides such an excellent fit to the $U-z$ magnitudes at these epochs
(Figure \ref{stritzinger.fig10}) that it must provide a better approximation to the true flux.
Figure \ref{stritzinger.fig12} (bottom) also reveals that the initial dip is less obvious
in the BB curve when we assume $E(B-V)$$_{host}$=0.00. Although this
hypothesis is very unlikely given the strong interstellar absorption lines
in the spectrum of SN~1999ex, this is illustrative of the effect reddening has on
the bolometric lightcurve of SN~1999ex, which will in turn affect future calculations of the progenitor system.

Among the three SN~Ib models of \citet{woosley87} model 6C 
-- characterized by a kinetic energy of 2.73$\times$10$^{51}$ ergs and 0.16 M$_\odot$ of $^{56}$Ni --
provides the best match to the BB lightcurve of SN~1999ex
(Figure \ref{stritzinger.fig12}, top panel). The agreement is remarkable
considering that we are not attempting to adjust the parameters.
The initial peak and subsequent dip have approximately the right luminosities
although the evolution of SN~1999ex was somewhat faster. The following
rise and post-maximum evolution is well described by the model. 

The observation of the tail of the shock wave breakout in SN~1999ex and the initial dip in
the lightcurve provides us with an insight on the
type of progenitor system for SNe~Ib/c. Several different models have been
proposed as progenitors for this type of SNe. One possibility is
an accreting white dwarf which may explode via thermal detonation 
upon reaching the Chandrasekhar mass \citep{sramek84,branch86}.
These models are expected to produce lightcurves with an initial peak that corresponds to the
emergence of the burning front, a fast luminosity drop due to adiabatic expansion,
and a subsequent rise caused by radioactive heating. Given the compact nature of
the progenitor ($\sim$1,800 km) the cooling time scale by adiabatic expansion
is only a few minutes (P. H\"oflich 2002, private communication) and the lightcurve
is entirely governed by radioactive heating \citep{hoflich96}. Hence,
these models are not expected to show an early dip at a few days past
explosion as is observed in SN~1999ex.
The second and more favored model for SNe~Ib/c consists of core collapse of
massive stars ($M_{ZAMS}$ $>$ 8 M$_{\odot}$) which lose their outer H envelope before
explosion. Within the core collapse models, there are two basic types
of progenitor systems: 1) a massive ($M_{ZAMS}$ $>$ 35 M$_{\odot}$) star which undergoes strong
stellar winds and becomes a Wolf-Rayet star at the time of explosion \citep{woosley93},
and 2) an exchanging binary system \citep{shigeyama90,nomoto94,iwamoto94} for less massive stars.
The resulting SNe have bolometric lightcurves containing an initial shock breakout followed by a dip.
Since the initial radii of these progenitors are $\sim$100 times greater than that of white
dwarfs, the dip occurs several days after explosion \citep{woosley87,shigeyama90,woosley93,woosley95}, very much like SN~1999ex. 


Although a detailed modeling of SN~1999ex is beyond the scope of this paper,
the fact that we were able to observe the initial shock break out followed by a dip within four
days since explosion lends support to the idea that SNe~Ib/c are due to the core collapse of 
massive progenitors rather than the thermonuclear disruption of white dwarfs.
This is a prediction that could not have been decisively made prior to these observations.   
         
\begin{figure}
\figurenum{13}
\epsscale{1.0}
\plotone{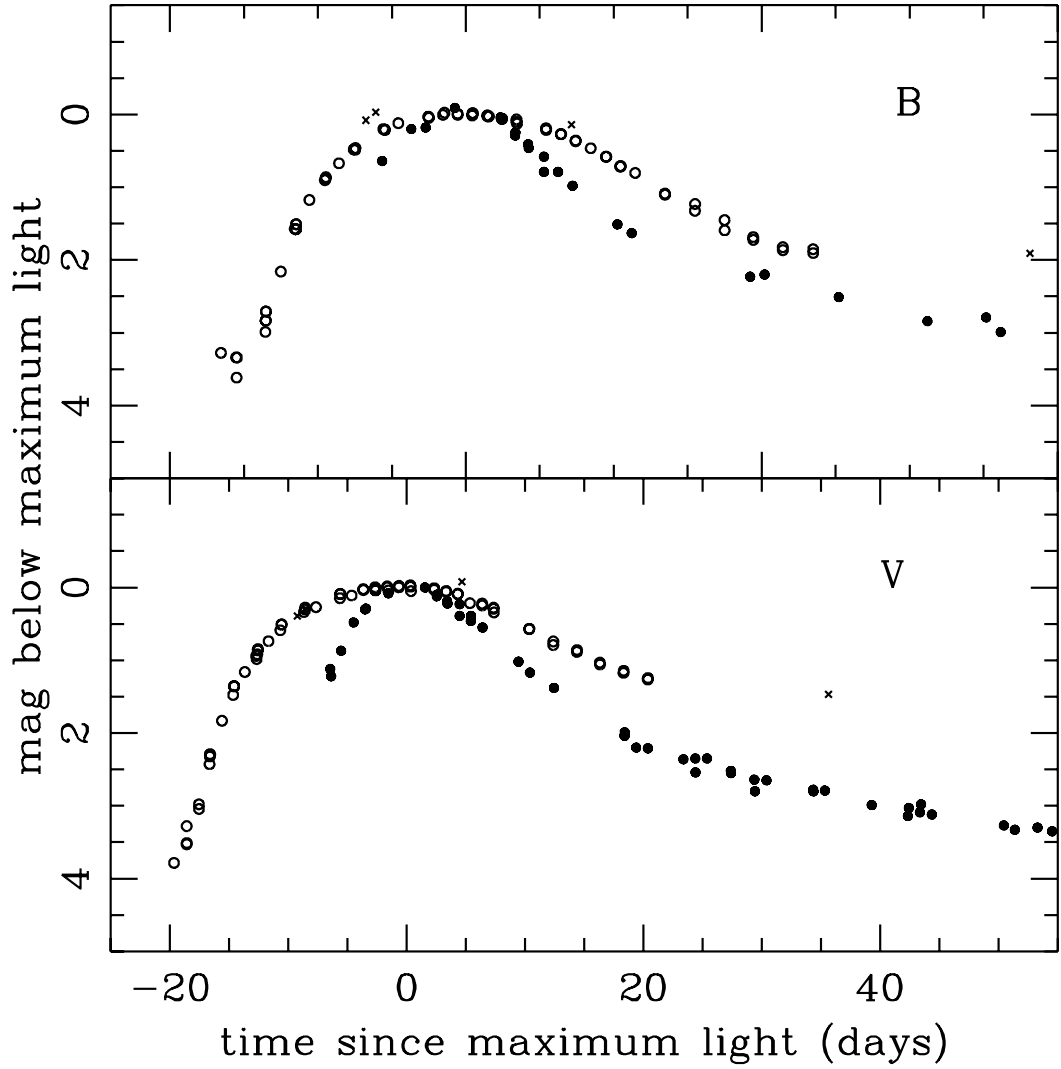}
\caption{Comparison of the $B$ and $V$-band lightcurves of SN~1999ex 
(open points), SN~1994I (closed points), and SN~1983V (crosses).
\label{stritzinger.fig13}}
\end{figure}

In Figure \ref{stritzinger.fig13} we present a comparison of the $B$ and $V$ band lightcurves of SN~1999ex
with SN~1994I \citep{richmond96} which proved spectroscopically similar to SN~1999ex \citep{hamuy02}
and the Type Ic SN~1983V \citep{clocchiatti97}.
Despite their photospheric similarities, Figure \ref{stritzinger.fig13} clearly shows that SN~1999ex
had a much broader peak compared to SN~1994I. 
In addition we see that after reaching maximum
light SN~1994I had a more rapid decline rate than SN~1999ex. These differences most likely
indicate that SN~1999ex was able to retain more of an envelope prior to core collapse 
thus, increasing the diffusion time for the energy produced from the radioactive 
decay of $^{56}Ni$ to $^{56}Co$ \citep{arnett96}.
SN~1983V on the other hand has very limited sampling of data, however the width of its 
lightcurve and subsequent tail are similar to SN~1999ex indicating a more
similar type of progenitor. 

\begin{figure}
\figurenum{14}
\epsscale{1.0}
\plotone{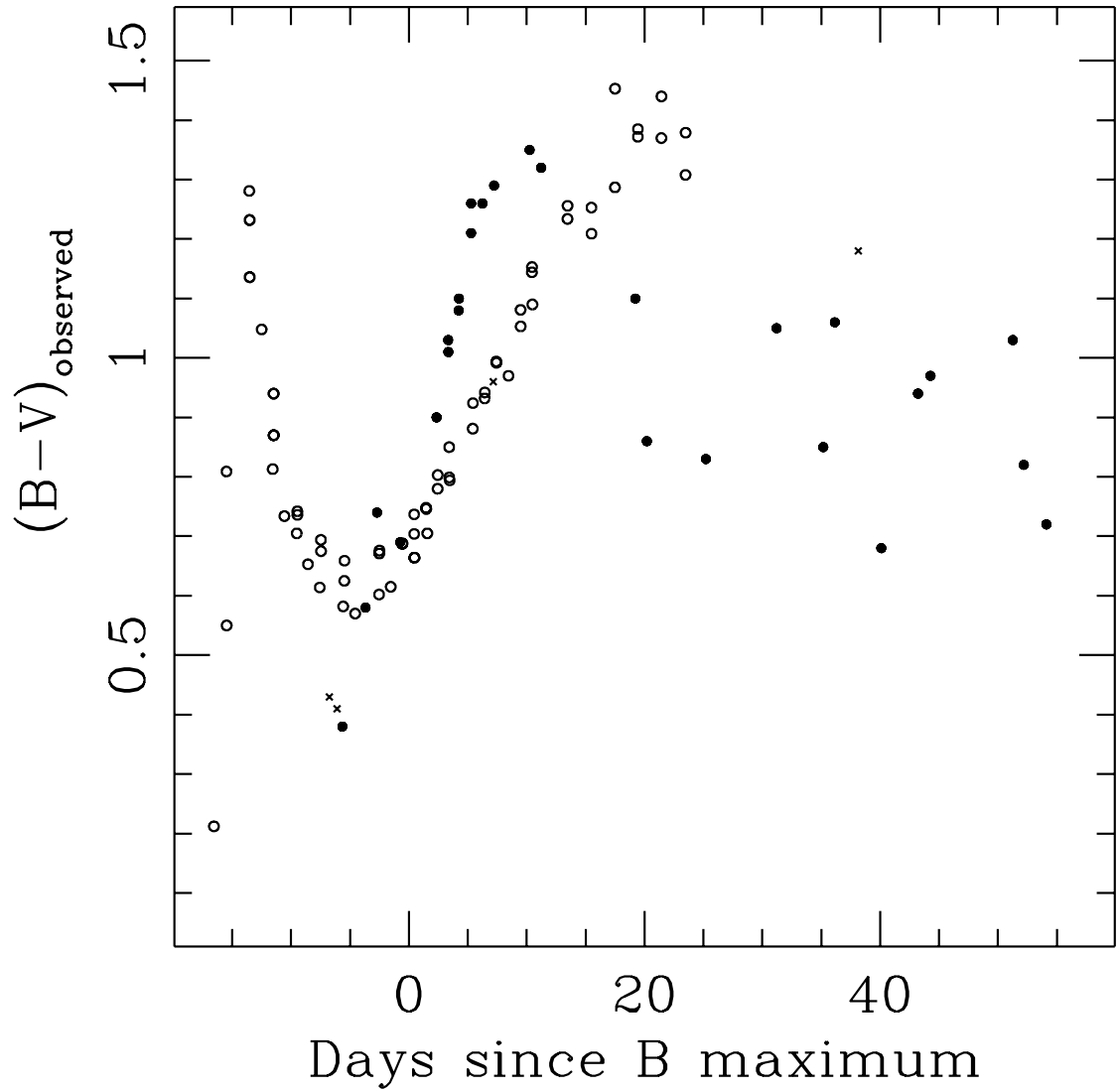}
\caption{Comparison of the $B-V$ color curves of SN~1999ex 
(open points), SN~1994I (closed points), and SN~1983V (crosses).
\label{stritzinger.fig14}}
\end{figure}

Figure \ref{stritzinger.fig14} displays the $B-V$ color curves of SN~1999ex, SN~1994I,
and SN~1983V.  From $B_{max}$ all three SNe evolved rapidly towards the red until approximately
15 to 20 days later in which they turned back towards the blue when the SNe became optically thin.
It is not possible to compare the color evolution at earlier epochs since SN~1999ex is the only object with 
data at this early time.



\subsection{Color evolution of Ia and Ib/c events}
\label{both.sec}

\begin{figure}
\figurenum{15}
\epsscale{1.0}
\plotone{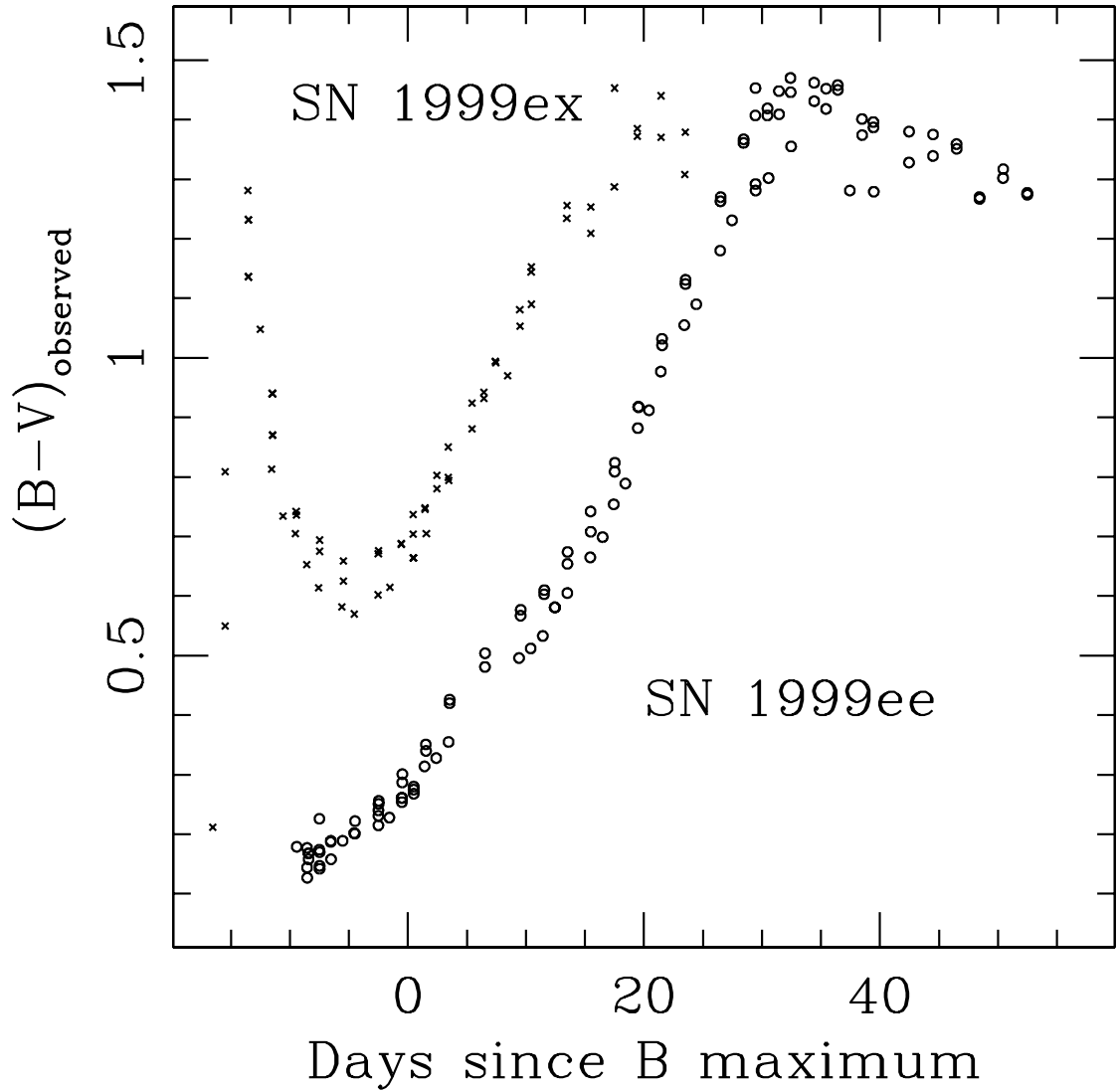}
\caption{$B-V$ color curves of SN~1999ee (open circles) and SN~1999ex (crosses) as a function of time since B maximum.
\label{stritzinger.fig15}}
\end{figure}

In the next years there will be several surveys of high-$z$ SNe, with the main
purpose of deriving cosmological parameters from SNe~Ia. Given the large
number of potential SNe to be discovered and the dim apparent magnitudes
of the high-$z$ objects, it will prove difficult to obtain a classification spectrum
for all of the SN candidates. Hence, it will prove important to figure out
photometric means to separate different types of SNe.
The lightcurves themselves will be valuable tools at separating SNe~II with
plateau like evolutions from the Ia's that display a bell-shaped lightcurve.
However, SNe~Ib/c and SNe~Ia have similar lightcurve shapes making
hard to distinguish both types. As shown in  Figure \ref{stritzinger.fig15}
the $B-V$ color curves of the Type Ia SN~1999ee and the Type Ib/c
SN~1999ex reveal very different behaviors, especially before maximum light.
But the post-maximum $B-V$ evolution of SN~1999ex is not so radically different than that of 
a fast-declining SNe~Ia. 
However, with early-time photometry, colors offer a potentially 
useful method to help discriminate between these two types of SNe.

\section{Conclusions}
\label{conc.sec}

In a rare occurrence two SNe were observed
simultaneously in IC~5179 at a redshift of z = 0.01 between 1999 October-December.
For each image of these SNe we performed galaxy subtraction of late time
images of their host galaxy after which we extracted $UBVRIz$ magnitudes.
We obtained well-sampled lightcurves including the first $z$ band observations
for a SN~Ia and SN~Ib/c.

Observations of the Type Ia SN~1999ee span from 10 days prior to $B_{max}$ to 53 days afterwards.
SN~1999ee was characterized by a slow decline rate of $\Delta$$m_{15}(B)$=0.94$\pm$0.06.
We estimate a value of host-galaxy reddening of 0.28$\pm$0.04 which we use
to derive reddening-free peak absolute magnitudes of $M_{B}$=-19.85$\pm$0.28,
$M_{V}$=-19.87$\pm$0.26, and $M_{I}$=-19.43$\pm$0.24. SN~1999ee is
among the most luminous SNe~Ia and fits well in the peak luminosity-decline rate
relation for SNe~Ia.

Near maximum we find systematic differences $\sim$0.05 mag
in photometry measured with two different telescopes, even though
the photometry is reduced to the same local standards around the supernova
using the specific color terms for each instrumental system. We use
models for our bandpasses and spectrophotometry of SN~1999ee
to derive magnitude corrections (S-corrections) and remedy this problem.
This exercise demonstrates the need of accurately characterizing the instrumental
system before great photometric accuracies of Type Ia supernovae can be claimed. It also shows
that this effect can  have important astrophysical consequences since a small
systematic shift of 0.02 mag in the $B-V$ color can   introduce a $0.08$ mag error
in the extinction corrected peak $B$ magnitudes of a supernova and thus
lead to biased cosmological parameters.

The data for the second SN observed in this galaxy -- Type~Ib/c SN~1999ex --
present us with the first ever observed shock breakout of a
supernova of this class. These observations show that shock breakout occurred
18 days before $B_{max}$ and support the idea that Type~Ib/c supernovae are due
to core collapse of massive stars rather than thermonuclear disruption of white dwarfs.
The comparison of the lightcurves of SN~1999ex to other SNe~Ic events like SN~1983V
and SN~1994I reveals large photometric differences among this class of objects,
probably due to variations in the properties of the envelopes of their progenitors
and/or explosion energies. Future theoretical modeling of this event
along with spectral analysis and construction of a bolometric lightcurve
will provide insight on relevant parameters describing its progenitor.

\acknowledgments
M.S. is very grateful to Cerro Tololo for allocating an office and 
providing computer facilities while in La Serena (where the first draft of this paper was prepared) 
as well as to Mark Wagner for providing computer facilities in Tucson. 
We are very grateful to the YALO team for their service observing observations
for this program and to the CTIO and ESO visitor support staffs for their
assistance in the course of our observing runs.
M.S. acknowledges support by the Hubble Space Telescope grant HST GO-07505.02-96A. 
M.H. acknowledges support provided by NASA through Hubble Fellowship grant HST-HF-01139.01-A
awarded by the Space Telescope Science Institute, which is operated by the Association
of Universities for Research in Astronomy, Inc., for NASA, under contract NAS 5-26555.
This research has made use of the NASA/IPAC Extragalactic Database (NED), which is operated by the
Jet Propulsion Laboratory, California Institute of Technology, under
contract with the National Aeronautics and Space Administration.
This research has made use of the SIMBAD database, operated at
CDS, Strasbourg, France.

\clearpage

\clearpage

\begin{deluxetable} {lccc}
\tablecolumns{4}
\tablenum{1}
\tablewidth{0pc}
\tablecaption{Journal of Observations \label{journal}}
\tablehead{
\colhead{Date(UT)} &
\colhead{Telescope} &
\colhead{Observatory} &
\colhead{Observer(s)} }

\startdata

1999 Oct. 08 & 0.91-m & Tololo   &Candia\\
1999 Oct. 09 & 0.91-m & Tololo   &Stubbs\\
1999 Oct. 10 & 0.91-m & Tololo   &Stubbs\\
1999 Oct. 10 & YALO   & Tololo   &Service Observing\\
1999 Oct. 11 & YALO   & Tololo   &Service Observing\\
1999 Oct. 11 & NTT    & La Silla &Hamuy/Doublier\\
1999 Oct. 11 & NTT    & La Silla &Hamuy/Doublier\\
1999 Oct. 12 & 0.91-m & Tololo   &Candia\\
1999 Oct. 13 & 0.91-m & Tololo   &Hamuy/P\'erez\\
1999 Oct. 13 & YALO   & Tololo   &Service Observing\\
1999 Oct. 15 & 0.91-m & Tololo   &Becker/Stubbs\\
1999 Oct. 15 & YALO   & Tololo   &Service Observing\\
1999 Oct. 16 & 0.91-m & Tololo   &Hamuy/Acevedo\\
1999 Oct. 17 & 0.91-m & Tololo   &Becker/Stubbs\\
1999 Oct. 17 & YALO   & Tololo   &Service Observing\\
1999 Oct. 18 & 0.91-m & Tololo   &Becker/Stubbs\\
1999 Oct. 19 & 0.91-m & Tololo   &Strolger\\
1999 Oct. 19 & YALO   & Tololo   &Service Observing\\
1999 Oct. 20 & 0.91-m & Tololo   &Strolger\\
1999 Oct. 20 & 1.54-m & La Silla &Pompei/Joguet/Sekiguchi\\
1999 Oct. 21 & 0.91-m & Tololo   &Strolger\\
1999 Oct. 21 & YALO   & Tololo   &Service Observing\\
1999 Oct. 24 & YALO   & Tololo   &Service Observing\\
1999 Oct. 25 & 1.54-m & La Silla &Augusteyn\\
1999 Oct. 27 & 0.91-m & Tololo   &Hamuy/Wischnjewsky\\
1999 Oct. 27 & YALO   & Tololo   &Service Observing\\
1999 Oct. 28 & 0.91-m & Tololo   &Wischnjewsky\\
1999 Oct. 29 & 1.54-m & La Silla &Augusteyn\\
1999 Oct. 29 & 0.91-m & Tololo   &Wischnjewsky\\
1999 Oct. 29 & YALO   & Tololo   &Service Observing\\
1999 Oct. 30 & 0.91-m & Tololo   &Wischnjewsky\\
1999 Oct. 31 & 0.91-m & Tololo   &Wischnjewsky\\
1999 Oct. 31 & YALO   & Tololo   &Service Observing\\
1999 Nov. 01 & 0.91-m & Tololo   &Wischnjewsky\\
1999 Nov. 02 & 0.91-m & Tololo   &Gonz\'alez\\
1999 Nov. 02 & YALO   & Tololo   &Service Observing\\
1999 Nov. 03 & 0.91-m & Tololo   &Gonz\'alez\\
1999 Nov. 04 & 0.91-m & Tololo   &Gonz\'alez\\
1999 Nov. 04 & YALO   & Tololo   &Service Observing\\
1999 Nov. 05 & 0.91-m & Tololo   &Gonz\'alez\\
1999 Nov. 06 & 0.91-m & Tololo   &Gonz\'alez\\
1999 Nov. 06 & YALO   & Tololo   &Service Observing\\
1999 Nov. 06 & 1.54-m & La Silla &Augusteyn\\
1999 Nov. 07 & 0.91-m & Tololo   &Gonz\'alez\\
1999 Nov. 08 & 0.91-m & Tololo   &Strolger\\
1999 Nov. 08 & YALO   & Tololo   &Service Observing\\
1999 Nov. 10 & 0.91-m & Tololo   &Strolger\\
1999 Nov. 10 & YALO   & Tololo   &Service Observing\\
1999 Nov. 11 & 0.91-m & Tololo   &Strolger\\
1999 Nov. 13 & 0.91-m & Tololo   &Rubenstein\\
1999 Nov. 13 & YALO   & Tololo   &Service Observing\\
1999 Nov. 14 & NTT    & La Silla &Hamuy/Doublier\\
1999 Nov. 14 & NTT    & La Silla &Hamuy/Doublier\\
1999 Nov. 14 & 0.91-m & Tololo   &Strolger\\
1999 Nov. 15 & YALO   & Tololo   &Service Observing\\
1999 Nov. 16 & 1.5-m  & Tololo   &Strolger\\
1999 Nov. 16 & YALO   & Tololo   &Service Observing\\
1999 Nov. 17 & 0.91-m & Tololo   &Rubenstein\\
1999 Nov. 17 & YALO   & Tololo   &Service Observing\\
1999 Nov. 18 & YALO   & Tololo   &Service Observing\\
1999 Nov. 19 & 0.91-m & Tololo   &Strolger\\
1999 Nov. 19 & YALO   & Tololo   &Service Observing\\
1999 Nov. 21 & YALO   & Tololo   &Service Observing\\
1999 Nov. 22 & YALO   & Tololo   &Service Observing\\
1999 Nov. 23 & YALO   & Tololo   &Service Observing\\
1999 Nov. 24 & 0.91-m & Tololo   &Hamuy/Antezana\\
1999 Nov. 25 & YALO   & Tololo   &Service Observing\\
1999 Nov. 26 & 0.91-m & Tololo   &Antezana\\
1999 Nov. 26 & YALO   & Tololo   &Service Observing\\
1999 Nov. 27 & 0.91-m & Tololo   &Antezana\\
1999 Nov. 29 & YALO   & Tololo   &Service Observing\\
1999 Dec. 01 & YALO   & Tololo   &Service Observing\\
1999 Dec. 03 & YALO   & Tololo   &Service Observing\\
1999 Dec. 05 & YALO   & Tololo   &Service Observing\\
1999 Dec. 07 & YALO   & Tololo   &Service Observing\\
1999 Dec. 09 & YALO   & Tololo   &Service Observing\\
2001 Jul. 16 & 1.5-m  & Tololo   &Candia\\
\enddata
\end{deluxetable}

\clearpage

\begin{deluxetable} {lcccccc}
\tablecolumns{7}
\tablenum{2}
\tablewidth{0pc}
\tablecaption{$UBVRIz$ Photometric Sequence Around IC~5179 \label{sequence}}
\tablehead{
\colhead{Star} &
\colhead{$U$} &
\colhead{$B$} &
\colhead{$V$} &
\colhead{$R$} &
\colhead{$I$} &
\colhead{$z$}  }

\startdata

c1  &14.282(012) & \nodata    & \nodata    & \nodata    & \nodata    &10.822(015)\\
c2  &13.663(010) & \nodata    & \nodata    & \nodata    & \nodata    &12.361(043)\\
c3  &19.579(043) &18.366(008) &16.866(009) &15.886(009) &14.793(009) &14.415(019)\\
c4  &17.183(012) &16.649(013) &15.789(011) &15.278(011) &14.826(011) &14.625(032)\\
c5  &18.985(035) &17.640(014) &16.334(015) &15.511(015) &14.817(015) &14.513(027)\\
c6  &20.532(163) &19.445(018) &18.009(009) &17.061(011) &16.048(009) &15.666(028)\\
c9  & \nodata    &21.366(240) &19.779(039) &18.830(062) &18.016(028) &17.574(057)\\
c11 &19.803(087) &20.033(046) &19.525(031) &19.234(038) &18.944(067) &19.287(228)\\
c12 &15.490(010) &15.418(008) &14.794(009) &14.407(011) &14.036(011) &13.880(025)\\
c13 &18.047(013) &18.161(011) &17.613(010) &17.246(009) &16.891(009) &16.675(044)\\
c14 &18.525(022) &18.701(010) &18.293(009) &18.011(009) &17.716(013) &17.545(014)\\
c15 &16.940(010) &16.618(008) &15.860(009) &15.401(009) &15.012(009) &14.864(011)\\
c16 &17.825(012) &17.695(008) &17.017(009) &16.616(009) &16.253(009) &16.087(024)\\
c17 & \nodata    &19.727(022) &18.233(011) &17.061(013) &15.537(011) &14.957(035)\\
c18 &14.761(015) &14.835(010) &14.346(011) &14.025(011) &13.707(011) &13.606(015)\\

\enddata
\end{deluxetable}

\clearpage

\begin{deluxetable} {lccccccc}
\tablecolumns{7}
\tablenum{3}
\tablewidth{0pc}
\tablecaption{Color Terms for the Different Photometric Systems \label{coefficients}}
\tablehead{
\colhead{Telescope} &
\colhead{$U$} &
\colhead{$B$} &
\colhead{$V$} &
\colhead{$R$} &
\colhead{$I$} &
\colhead{$z$}  }

\startdata

CTIO 0.9-m  & +0.140 & -0.083 & +0.017 & +0.019 & +0.020  & +0.007 \\ 
YALO        & +0.079 & +0.098 & -0.019 & +0.352 & -0.020  & \nodata \\
CTIO 1.54-m & +0.139 & -0.080 & +0.030 & +0.016 & +0.017  & \nodata \\
NTT         & +0.061 & -0.121 & +0.020 & +0.007 & -0.035  & 0.000   \\
D1.54-m     & \nodata & \nodata & +0.033 & \nodata & \nodata & \nodata \\

\enddata
\end{deluxetable}

\clearpage

\begin{deluxetable} {lcccccccc}
\rotate
\tablecolumns{9}
\tablenum{4}
\tablewidth{0pc}
\tablecaption{$UBVRIz$ Photometry for SN~1999ee \label{eephotometry}}
\tablehead{
\colhead{JD-2,451,000} &
\colhead{$U$} &
\colhead{$B$} &
\colhead{$V$} &
\colhead{$R$} &
\colhead{$I$} &
\colhead{$z$} &
\colhead{Method} &
\colhead{Telescope} }
\startdata

459.68 & \nodata     & 15.808(014) & 15.629(015) & 15.410(015) & 15.354(015) & \nodata      & PHOT  & CTIO 0.9-m \\
460.56 & 15.440(017) & 15.575(014) & 15.431(015) & 15.221(015) & 15.167(015) & \nodata      & PHOT  & CTIO 0.9-m \\
460.56 & 15.462(017) & 15.596(014) & 15.419(015) & 15.230(015) & 15.166(015) & \nodata      & PHOT  & CTIO 0.9-m \\
460.57 & 15.473(017) & 15.585(014) & 15.458(015) & 15.210(015) & 15.191(015) & \nodata      & PHOT  & CTIO 0.9-m \\
460.68 & 15.419(017) & 15.571(014) & 15.403(015) & 15.203(015) & 15.149(015) & \nodata      & PHOT  & CTIO 0.9-m \\
460.68 & 15.437(017) & 15.570(014) & 15.412(015) & 15.197(015) & 15.146(015) & \nodata      & PHOT  & CTIO 0.9-m \\
460.68 & 15.422(017) & 15.578(014) & 15.410(015) & 15.200(015) & 15.137(015) & \nodata      & PHOT  & CTIO 0.9-m \\
461.59 & 15.249(016) & 15.394(017) & 15.220(015) & 15.019(011) & 14.937(015) & \nodata      & DAO   & YALO \\
461.60 & 15.212(016) & 15.431(017) & 15.205(015) & 15.002(011) & 14.942(015) & \nodata      & DAO   & YALO \\
461.62 & 15.265(082) & 15.413(014) & 15.271(015) & 15.033(015) & 14.984(015) & \nodata      & PHOT  & CTIO 0.9-m \\
461.62 & 15.247(017) & 15.425(014) & 15.255(015) & 15.034(015) & 14.976(015) & \nodata      & PHOT  & CTIO 0.9-m \\
461.62 & 15.256(021) & 15.392(014) & 15.245(015) & 15.044(015) & 14.983(015) & \nodata      & PHOT  & CTIO 0.9-m \\
462.56 & 15.033(016) & 15.274(017) & 15.087(015) & 14.908(011) & 14.820(015) & \nodata      & DAO   & YALO \\
462.57 & 15.006(016) & 15.279(017) & 15.090(015) & 14.904(011) & 14.812(015) & \nodata      & DAO   & YALO \\
462.59 & 15.195(016) & 15.264(013) & 15.106(015) & 14.907(015) & 14.878(016) & 14.769(015)  & PHOT  & NTT \\
463.57 & 14.977(017) & 15.171(014) & 14.982(015) & 14.777(015) & 14.763(015) & \nodata      & PHOT  & CTIO 0.9-m \\
464.55 & 14.892(017) & 15.090(014) & 14.888(015) & 14.712(015) & 14.705(015) & \nodata      & PHOT  & CTIO 0.9-m \\
464.58 & \nodata     & \nodata     & \nodata     & 14.703(015) & 14.717(015) & \nodata      & PHOT  & CTIO 0.9-m \\
464.61 & 14.758(016) & 15.068(017) & 14.867(015) & 14.703(011) & 14.674(015) & \nodata      & PHOT  & YALO \\
464.62 & 14.782(016) & 15.083(017) & 14.861(015) & 14.699(011) & 14.681(015) & \nodata      & PHOT  & YALO \\
466.60 & \nodata     & 14.981(014) & 14.750(015) & 14.600(015) & \nodata     & \nodata      & PHOT  & CTIO 0.9-m \\
466.60 & \nodata     & 14.965(014) & 14.750(015) & 14.605(015) & \nodata     & \nodata      & PHOT  & CTIO 0.9-m \\
466.60 & \nodata     & 14.976(014) & 14.736(015) & 14.605(015) & \nodata     & \nodata      & PHOT  & CTIO 0.9-m \\
466.62 & 14.685(016) & 14.969(017) & 14.718(015) & 14.607(011) & 14.651(015) & \nodata      & PHOT  & YALO \\
466.63 & 14.664(016) & 14.970(017) & 14.714(015) & 14.607(011) & 14.670(015) & \nodata      & PHOT  & YALO \\
467.53 & 14.807(017) & 14.937(014) & 14.709(015) & 14.567(015) & 14.670(015) & 14.517(015)  & PHOT  & CTIO 0.9-m \\
467.54 & 14.802(017) & \nodata     & \nodata     & \nodata     & \nodata     & \nodata      & PHOT  & CTIO 0.9-m \\
467.54 & 14.804(017) & \nodata     & \nodata     & \nodata     & \nodata     & \nodata      & PHOT  & CTIO 0.9-m \\
468.59 & \nodata     & 14.934(014) & 14.680(015) & 14.544(015) & \nodata     & \nodata      & PHOT  & CTIO 0.9-m \\
468.59 & \nodata     & 14.940(014) & 14.679(015) & 14.545(015) & \nodata     & \nodata      & PHOT  & CTIO 0.9-m \\
468.59 & \nodata     & 14.942(014) & 14.681(015) & 14.553(015) & \nodata     & \nodata      & PHOT  & CTIO 0.9-m \\
468.62 & 14.683(018) & 14.931(017) & 14.644(015) & 14.563(011) & 14.680(015) & \nodata      & DAO   & YALO \\
468.63 & 14.687(016) & 14.938(017) & 14.637(015) & 14.561(011) & 14.666(015) & \nodata      & DAO   & YALO \\
469.59 & \nodata     & 14.941(014) & 14.661(015) & 14.517(015) & \nodata     & \nodata      & PHOT  & CTIO 0.9-m \\
469.60 & \nodata     & 14.926(014) & 14.658(015) & 14.523(015) & \nodata     & \nodata      & PHOT  & CTIO 0.9-m \\
469.60 & \nodata     & 14.938(014) & 14.663(015) & 14.520(015) & \nodata     & \nodata      & PHOT  & CTIO 0.9-m \\
470.51 & 14.899(017) & 14.958(014) & 14.644(015) & 14.505(015) & 14.733(015) & \nodata      & PHOT  & CTIO 0.9-m \\
470.62 & 14.826(016) & 14.948(017) & 14.597(015) & 14.535(011) & 14.721(015) & \nodata      & DAO   & YALO \\
470.63 & 14.801(032) & 14.953(017) & 14.613(015) & 14.544(011) & 14.733(015) & \nodata      & DAO   & YALO \\
471.50 & 14.962(017) & 14.968(014) & 14.640(015) & 14.498(015) & 14.765(015) & \nodata      & PHOT  & CTIO 0.9-m \\
471.54 & \nodata     & \nodata     & 14.653(015) & \nodata     & \nodata     & \nodata      & PHOT  & D 1.54-m \\
471.55 & \nodata     & \nodata     & 14.652(015) & \nodata     & \nodata     & \nodata      & PHOT  & D 1.54-m \\ 
471.55 & \nodata     & \nodata     & 14.637(015) & \nodata     & \nodata     & \nodata      & PHOT  & D 1.54-m \\
471.59 & \nodata     & \nodata     & 14.647(015) & \nodata     & \nodata     & \nodata      & PHOT  & D 1.54-m \\
472.55 & 15.036(017) & 15.001(014) & 14.646(015) & 14.513(015) & 14.797(015) & \nodata      & PHOT  & CTIO 0.9-m \\
472.63 & 14.977(016) & 15.032(017) & 14.612(015) & 14.569(011) & 14.810(015) & \nodata      & DAO   & YALO \\
472.64 & 14.988(018) & 15.032(017) & 14.606(015) & 14.569(011) & 14.811(015) & \nodata      & DAO   & YALO \\
475.63 & 15.281(018) & 15.155(017) & 14.674(015) & 14.645(011) & 14.933(015) & \nodata      & DAO   & YALO \\
475.64 & 15.284(017) & 15.165(017) & 14.661(015) & 14.642(011) & 14.928(015) & \nodata      & DAO   & YALO \\
476.50 & \nodata     & \nodata     & 14.759(015) & \nodata     & \nodata     & \nodata      & PHOT  & D 1.54-m \\
476.51 & \nodata     & \nodata     & 14.753(015) & \nodata     & \nodata     & \nodata      & PHOT  & D 1.54-m \\
478.52 & 15.536(017) & 15.335(014) & 14.839(015) & 14.738(015) & 15.116(015) & 14.746(015)  & PHOT  & CTIO 0.9-m  \\
478.65 & 15.628(019) & 15.369(017) & 14.802(015) & 14.797(011) & 15.120(015) & \nodata      & DAO   & YALO \\
478.66 & 15.638(021) & 15.365(017) & 14.788(015) & 14.788(011) & 15.111(015) & \nodata      & DAO   & YALO \\
479.51 & 15.646(017) & 15.405(014) & 14.893(015) & 14.810(015) & 15.173(015) & 14.787(015)  & PHOT  & CTIO 0.9-m \\
480.50 & \nodata     & \nodata     & 14.981(015) & \nodata     & \nodata     & \nodata      & PHOT  & D 1.54-m \\
480.53 & 15.784(017) & 15.486(014) & 14.953(015) & 14.877(015) & 15.245(015) & 14.813(015)  & PHOT  & CTIO 0.9-m \\
480.64 & 16.003(027) & 15.521(017) & 14.918(015) & 14.934(011) & 15.236(015) & \nodata      & DAO   & YALO \\
480.66 & 15.943(096) & 15.528(017) & 14.918(015) & 14.931(011) & 15.267(015) & \nodata      & DAO   & YALO \\
481.55 & 15.918(017) & 15.581(014) & 15.000(015) & 14.932(015) & 15.295(015) & 14.837(015)  & DAO   & CTIO 0.9-m \\
481.55 & 15.918(017) & 15.581(014) & 15.000(015) & 14.932(015) & 15.295(015) & 14.837(015)  & DAO   & CTIO 0.9-m \\
482.58 & 16.047(017) & 15.709(014) & \nodata     & \nodata     & \nodata     & \nodata      & DAO   & CTIO 0.9-m \\
482.61 & 16.078(017) & 15.715(014) & 15.110(015) & 15.050(015) & 15.362(015) & 14.833(015)  & DAO   & CTIO 0.9-m \\
482.62 & 16.366(051) & 15.703(017) & 15.049(015) & 15.047(011) & 15.386(015) & \nodata      & DAO   & YALO \\
482.63 & 16.377(090) & 15.722(017) & 15.048(015) & 15.052(011) & 15.365(015) & \nodata      & DAO   & YALO \\
483.65 & 16.207(017) & \nodata     & 15.154(015) & 15.097(015) & 15.365(015) & 14.834(015)  & DAO   & CTIO 0.9-m \\
483.66 & \nodata     & \nodata     & 15.156(015) & \nodata     & \nodata     & \nodata      & DAO   & CTIO 0.9-m \\
484.55 & 16.365(017) & 15.889(014) & 15.224(015) & 15.135(015) & 15.396(015) & 14.836(015)  & DAO   & CTIO 0.9-m \\
484.58 & 16.617(040) & 15.926(017) & 15.184(015) & 15.147(011) & 15.427(015) & \nodata      & DAO   & YALO \\
484.59 & \nodata     & 15.885(017) & 15.177(015) & 15.143(011) & 15.427(015) & \nodata      & DAO   & YALO \\
485.60 & 16.507(017) & 15.995(014) & 15.296(015) & 15.190(015) & 15.410(015) & 14.803(015)  & DAO   & CTIO 0.9-m \\
486.54 & 16.648(017) & 16.098(014) & 15.344(015) & 15.217(015) & 15.401(015) & 14.798(015)  & DAO   & CTIO 0.9-m \\
486.62 & 16.902(050) & 16.105(017) & 15.296(015) & 15.212(011) & 15.438(015) & \nodata      & DAO   & YALO \\
486.63 & 16.982(066) & 16.107(017) & 15.283(015) & 15.208(011) & 15.425(015) & \nodata      & DAO   & YALO \\
487.53 & 16.768(017) & 16.195(014) & 15.406(015) & 15.257(015) & 15.381(015) & 14.774(015)  & DAO   & CTIO 0.9-m \\
488.50 & \nodata     & \nodata     & 15.456(015) & \nodata     & \nodata     & \nodata      & DAO   & D 1.54-m \\
488.53 & \nodata     & \nodata     & 15.442(015) & \nodata     & \nodata     & \nodata      & PHOT  & D 1.54-m \\
488.58 & 16.891(017) & 16.293(014) & 15.411(015) & 15.240(015) & 15.351(015) & 14.761(015)  & DAO   & CTIO 0.9-m \\
488.63 & 17.131(051) & 16.300(017) & 15.382(015) & 15.241(011) & 15.402(015) & \nodata      & DAO   & YALO \\
488.64 & 17.144(056) & 16.304(017) & 15.387(015) & 15.243(011) & 15.394(015) & \nodata      & DAO   & YALO \\
489.53 & 17.006(017) & 16.372(014) & 15.460(015) & 15.255(018) & 15.351(015) & 14.761(015)  & DAO   & CTIO 0.9-m \\
490.53 & 17.126(017) & 16.468(014) & 15.491(015) & 15.261(016) & 15.315(015) & \nodata      & DAO   & CTIO 0.9-m \\
490.63 & 17.330(054) & 16.501(017) & 15.469(015) & 15.273(011) & 15.340(015) & \nodata      & DAO   & YALO \\
490.64 & 17.344(067) & 16.488(017) & 15.467(015) & 15.273(011) & 15.355(015) & \nodata      & DAO   & YALO \\
492.52 & 17.319(017) & 16.641(014) & 15.586(015) & 15.284(015) & 15.274(015) & \nodata      & DAO   & CTIO 0.9-m \\
492.62 & 17.356(058) & 16.664(017) & 15.540(015) & 15.299(011) & 15.289(015) & \nodata      & DAO   & YALO \\
492.63 & 17.427(094) & 16.668(017) & 15.537(015) & 15.297(011) & 15.295(015) & \nodata      & DAO   & YALO \\
493.54 & 17.421(017) & 16.722(014) & 15.632(015) & 15.301(015) & 15.264(015) & \nodata      & DAO   & CTIO 0.9-m \\
495.57 & 17.570(021) & 16.898(014) & 15.718(015) & 15.334(015) & 15.235(015) & \nodata      & DAO   & CTIO 0.9-m \\
495.59 & 17.781(079) & 16.925(017) & 15.662(015) & 15.345(011) & 15.247(015) & \nodata      & DAO   & YALO \\
495.60 & 17.807(106) & 16.935(017) & 15.665(015) & 15.347(011) & 15.239(015) & \nodata      & DAO   & YALO \\
496.56 & 17.672(017) & 16.975(014) & 15.744(015) & 15.321(015) & 15.168(015) & \nodata      & DAO   & CTIO 0.9-m \\
496.66 & \nodata     & 16.952(013) & \nodata     & 15.330(015) & \nodata     & \nodata      & PHOT  & NTT \\
497.54 & 18.008(154) & 17.094(017) & 15.733(015) & 15.379(011) & 15.196(015) & \nodata      & DAO   & YALO \\
497.55 & 17.837(095) & 17.109(017) & 15.742(015) & 15.380(011) & 15.211(015) & \nodata      & DAO   & YALO \\
498.54 & 17.866(017) & \nodata     & \nodata     & \nodata     & \nodata     & \nodata      & PHOT  & CTIO 1.5-m \\
498.54 & 17.954(109) & 17.180(017) & 15.773(015) & 15.398(011) & 15.186(015) & \nodata      & DAO   & YALO \\
498.55 & 17.897(120) & 17.231(017) & 15.778(015) & 15.404(011) & 15.185(015) & \nodata      & DAO   & YALO \\
498.56 & 17.790(024) & 17.122(014) & 15.830(015) & 15.381(015) & 15.154(015) & \nodata      & PHOT  & CTIO 1.5-m \\
498.57 & \nodata     & 17.114(014) & 15.833(015) & 15.386(015) & 15.159(015) & \nodata      & PHOT  & CTIO 1.5-m \\
499.56 & 18.089(090) & 17.235(017) & 15.828(015) & 15.446(011) & 15.178(015) & \nodata      & DAO   & YALO \\
499.57 & 18.018(107) & 17.243(017) & 15.824(015) & 15.442(011) & 15.181(015) & \nodata      & DAO   & YALO \\
499.66 & 17.852(023) & 17.186(014) & 15.884(015) & 15.408(015) & 15.152(015) & \nodata      & DAO   & CTIO 0.9-m \\
500.54 & 17.949(093) & 17.323(017) & 15.875(015) & 15.464(011) & 15.193(015) & \nodata      & DAO   & YALO \\
500.55 & 18.112(131) & 17.300(017) & 15.891(015) & 15.476(011) & 15.206(015) & \nodata      & DAO   & YALO \\
501.52 & 17.988(134) & 17.397(017) & 15.927(015) & 15.517(011) & 15.207(015) & \nodata      & DAO   & YALO \\
501.53 & 18.103(128) & 17.387(017) & 15.941(015) & 15.520(011) & 15.202(015) & \nodata      & DAO   & YALO \\
501.55 & 18.037(030) & \nodata     & \nodata     & \nodata     & \nodata     & \nodata      & DAO   & CTIO 0.9-m \\
501.57 & 17.976(031) & 17.352(014) & 15.997(015) & 15.494(015) & 15.161(015) & 14.728(015)  & DAO   & CTIO 0.9-m \\
503.52 & 18.115(172) & 17.525(017) & 16.063(015) & 15.624(011) & 15.272(015) & \nodata      & DAO   & YALO \\
503.53 & 18.100(152) & 17.495(017) & 16.064(015) & 15.638(011) & 15.274(015) & \nodata      & DAO   & YALO \\
504.53 & 18.093(147) & 17.566(017) & 16.148(015) & 15.709(011) & 15.358(015) & \nodata      & DAO   & YALO \\
504.54 & \nodata     & 17.587(017) & 16.135(015) & 15.697(011) & 15.335(015) & \nodata      & DAO   & YALO \\
505.52 & \nodata     & 17.671(018) & 16.213(015) & 15.781(011) & 15.423(015) & \nodata      & DAO   & YALO \\
505.53 & 18.361(155) & 17.644(018) & 16.194(015) & 15.771(011) & 15.417(015) & \nodata      & DAO   & YALO \\
506.55 & 18.237(040) & 17.611(014) & 16.330(015) & 15.837(017) & 15.428(015) & 14.963(016)  & DAO   & CTIO 0.9-m \\
507.58 & \nodata     & 17.724(020) & 16.323(015) & 15.925(011) & 15.570(015) & \nodata      & DAO   & YALO \\
507.59 & \nodata     & 17.731(023) & 16.357(015) & 15.925(011) & 15.571(015) & \nodata      & DAO   & YALO \\
508.54 & 18.302(113) & 17.777(017) & 16.381(015) & 15.978(011) & 15.620(015) & \nodata      & DAO   & YALO \\
508.55 & \nodata     & 17.772(017) & 16.385(015) & 15.982(011) & 15.627(015) & \nodata      & DAO   & YALO \\
508.58 & 18.331(030) & 17.712(014) & 16.433(015) & 15.954(015) & 15.581(015) & 15.134(015)  & DAO   & CTIO 0.9-m \\
509.55 & 18.333(027) & 17.749(014) & \nodata     & 16.008(015) & 15.644(015) & 15.192(015)  & DAO   & CTIO 0.9-m \\
511.56 & \nodata     & 17.857(017) & 16.529(015) & 16.136(011) & 15.825(015) & \nodata      & DAO   & YALO \\
511.57 & \nodata     & 17.892(018) & 16.512(015) & 16.140(011) & 15.807(015) & \nodata      & DAO   & YALO \\
513.59 & \nodata     & 17.956(031) & 16.617(015) & 16.257(011) & 15.955(015) & \nodata      & DAO   & YALO \\
513.60 & \nodata     & 18.005(031) & 16.630(015) & 16.250(011) & 15.963(015) & \nodata      & DAO   & YALO \\
515.59 & \nodata     & 18.026(030) & 16.667(015) & 16.317(011) & 16.044(015) & \nodata      & DAO   & YALO \\
515.60 & \nodata     & 18.012(022) & 16.661(015) & 16.307(011) & 16.006(015) & \nodata      & DAO   & YALO \\
517.53 & \nodata     & 18.002(019) & 16.732(015) & 16.395(011) & 16.106(015) & \nodata      & DAO   & YALO \\
517.54 & 18.906(181) & 18.004(020) & 16.737(015) & 16.398(011) & 16.128(015) & \nodata      & DAO   & YALO \\
519.53 & \nodata     & 18.076(022) & 16.774(015) & 16.467(011) & 16.212(015) & \nodata      & DAO   & YALO \\
519.54 & \nodata     & 18.092(023) & 16.775(015) & 16.472(011) & 16.206(015) & \nodata      & DAO   & YALO \\
521.58 & \nodata     & 18.110(021) & 16.833(015) & 16.534(011) & 16.287(015) & \nodata      & DAO   & YALO \\
521.59 & \nodata     & 18.106(021) & 16.832(015) & 16.535(011) & 16.266(015) & \nodata      & DAO   & YALO \\

\enddata
\tablecomments{DAO means PSF photometry, while PHOT means aperture photometry}
\end{deluxetable}

\clearpage

\begin{deluxetable} {lcccccc}
\tablecolumns{6}
\tablenum{5}
\tablewidth{0pc}
\tablecaption{Wavelength Shifts to Instrumental Bandpasses \label{shifts}}
\tablehead{
\colhead{} &
\colhead{$B$} &
\colhead{$V$} &
\colhead{$R$} &
\colhead{$I$} &
\colhead{$z$} \\
\colhead{Telescope} &
\colhead{(\AA)} &
\colhead{(\AA)} &
\colhead{(\AA)} &
\colhead{(\AA)} &
\colhead{(\AA)} }

\startdata

CTIO 0.9-m  &    0    &  15 blue &  30 blue & 90 red  & 50 blue \\ 
YALO        &  30 red & 120 blue & 180 blue & 10 blue & \nodata \\

\enddata
\end{deluxetable}

\clearpage

\begin{deluxetable} {lcc}
\tablecolumns{3}
\tablenum{6}
\tablewidth{0pc}
\tablecaption{Lightcurve Parameters  \label{lcpar}}
\tablehead{
\colhead{Filter} &
\colhead{Peak Mag.} &
\colhead{JD Max} }
\startdata

&SN~1999ee&\\
$U$ &14.74(0.06) &467.5(0.5)\\
$B$ &14.93(0.02) &469.1(0.5)\\
$V$ &14.61(0.03) &471.7(0.5)\\
$R$ &14.53(0.02) &470.8(0.5)\\
$I$ &14.66(0.02) &466.3(0.5)\\
$z$ &14.50(0.05) &468.5(1.0)\\
&SN~1999ex&\\
$U$ &17.43(0.12) &496.2(0.5)\\
$B$ &17.35(0.02) &498.1(0.5)\\
$V$ &16.63(0.04) &501.2(0.5)\\
$R$ &16.26(0.02) &502.7(0.5)\\
$I$ &15.98(0.02) &502.6(0.5)\\
$z$ &15.82(0.05) &504.2(1.0)\\

\enddata
\end{deluxetable}

\clearpage

\begin{deluxetable} {lcccccccc}
\rotate
\tablecolumns{9}
\tablenum{7}
\tablewidth{0pc}
\tablecaption{$UBVRIz$ Photometry for SN~1999ex \label{exphotometry}}
\tablehead{
\colhead{JD-2,451,000} &
\colhead{$U$} &
\colhead{$B$} &
\colhead{$V$} &
\colhead{$R$} &
\colhead{$I$} &
\colhead{$z$} &
\colhead{Method} &
\colhead{Telescope} }
\startdata

481.55 & 19.876(108) & 20.626(058) & 20.414(057) & 20.164(064) & 19.816(101) & 19.946(208) & DAO  & CTIO 0.91-m \\
482.58 & 20.687(204) & 20.689(109) & \nodata     & \nodata     & \nodata     & \nodata     & DAO  & CTIO 0.91-m \\
482.61 & 20.500(163) & 20.966(116) & 20.157(079) & 19.360(056) & 19.592(147) & 19.471(219) & DAO  & CTIO 0.91-m \\
482.62 & \nodata     & \nodata     & 19.909(166) & 19.489(080) & 19.328(247) & 0000000000) & DAO  & YALO \\
482.63 & \nodata     & 20.691(222) & 20.141(198) & 19.559(087) & 19.316(314) & 0000000000) & DAO  & YALO \\
483.65 & \nodata     & \nodata     & 19.609(039) & 19.167(044) & 18.990(051) & 18.668(071) & DAO  & CTIO 0.91-m \\
483.66 & \nodata     & \nodata     & 19.673(036) & \nodata     & \nodata     & \nodata     & DAO  & CTIO 0.91-m \\
484.55 & 20.981(173) & 20.339(055) & 19.058(019) & 18.608(023) & 18.510(033) & 18.165(034) & DAO  & CTIO 0.91-m \\
484.58 & \nodata     & 20.183(119) & 18.951(043) & 18.569(023) & 18.344(068) & \nodata     & DAO  & YALO \\
484.58 & \nodata     & 20.183(119) & 18.951(043) & 18.569(023) & 18.344(068) & \nodata     & DAO  & YALO \\
484.59 & \nodata     & 20.058(107) & 18.922(054) & 18.533(027) & 18.388(064) & \nodata     & DAO  & YALO \\
484.59 & \nodata     & 20.058(107) & 18.922(054) & 18.533(027) & 18.388(064) & \nodata     & DAO  & YALO \\
485.60 & 20.040(132) & 19.509(029) & 18.461(019) & 18.014(024) & 17.969(034) & 17.800(051) & DAO  & CTIO 0.91-m \\
486.54 & 19.248(041) & 18.920(017) & 18.107(015) & 17.715(017) & 17.633(017) & 17.417(025) & DAO  & CTIO 0.91-m \\
486.62 & \nodata     & 18.928(044) & 17.988(028) & 17.676(015) & 17.547(040) & \nodata     & DAO  & YALO \\
486.62 & \nodata     & 18.928(044) & 17.988(028) & 17.676(015) & 17.547(040) & \nodata     & DAO  & YALO \\
486.63 & \nodata     & 18.858(042) & 17.988(025) & 17.666(015) & 17.601(041) & \nodata     & DAO  & YALO \\
486.63 & \nodata     & 18.858(042) & 17.988(025) & 17.666(015) & 17.601(041) & \nodata     & DAO  & YALO \\
487.53 & 18.788(033) & 18.525(014) & 17.791(015) & 17.431(015) & 17.354(019) & 17.101(021) & DAO  & CTIO 0.91-m \\
488.50 & \nodata     & \nodata     & 17.570(027) & \nodata     & \nodata     & \nodata     & DAO  & D 1.54-m \\
488.53 & \nodata     & \nodata     & 17.613(020) & \nodata     & \nodata     & \nodata     & PHOT & D 1.54-m \\
488.58 & 18.422(029) & 18.252(015) & 17.547(017) & 17.254(016) & 17.161(015) & 16.884(019) & DAO  & CTIO 0.91-m \\
488.63 & 18.359(145) & 18.236(024) & 17.494(017) & 17.200(011) & 17.072(022) & \nodata     & DAO  & YALO \\
488.64 & 18.307(150) & 18.212(024) & 17.476(015) & 17.194(011) & 17.004(018) & \nodata     & DAO  & YALO \\
489.53 & 18.196(030) & 18.021(014) & 17.368(015) & 17.077(017) & 16.943(017) & 16.707(018) & DAO  & CTIO 0.91-m \\
490.53 & 17.975(022) & 17.831(014) & 17.217(015) & 16.915(017) & 16.801(016) & \nodata     & DAO  & CTIO 0.91-m \\
490.63 & 17.762(094) & 17.835(017) & 17.141(015) & 16.859(011) & 16.735(017) & \nodata     & DAO  & YALO \\
490.64 & 17.686(083) & 17.811(017) & 17.136(015) & 16.863(011) & 16.748(017) & \nodata     & DAO  & YALO \\
492.52 & 17.679(018) & 17.550(014) & 16.968(015) & 16.683(015) & 16.514(015) & \nodata     & DAO  & CTIO 0.91-m \\
492.62 & 17.426(069) & 17.556(017) & 16.931(015) & 16.643(011) & 16.490(015) & \nodata     & DAO  & YALO \\
492.63 & 17.783(135) & 17.565(017) & 16.906(015) & 16.642(011) & 16.488(015) & \nodata     & DAO  & YALO \\
493.54 & 17.563(017) & 17.470(014) & 16.900(015) & 16.585(015) & 16.427(015) & \nodata     & DAO  & CTIO 0.91-m \\
495.57 & 17.460(017) & 17.379(014) & 16.777(015) & 16.460(015) & 16.278(016) & \nodata     & DAO  & CTIO 0.91-m \\
495.59 & 17.304(057) & 17.391(017) & 16.720(015) & 16.424(011) & 16.233(015) & \nodata     & DAO  & YALO \\
495.60 & 17.265(064) & 17.395(017) & 16.719(015) & 16.412(011) & 16.222(015) & \nodata     & DAO  & YALO \\
496.56 & 17.503(017) & 17.355(014) & 16.740(015) & 16.393(015) & 16.185(015) & \nodata     & DAO  & CTIO 0.91-m \\
496.66 & \nodata     & 17.326(013) & \nodata     & 16.407(015) & \nodata     & \nodata     & PHOT & NTT \\
497.54 & 17.486(108) & 17.340(019) & 16.653(015) & 16.329(011) & 16.094(015) & \nodata     & DAO  & YALO \\
497.55 & 17.371(058) & 17.353(017) & 16.665(015) & 16.335(011) & 16.116(015) & \nodata     & DAO  & YALO \\
498.54 & 17.520(070) & 17.364(017) & 16.627(015) & 16.298(011) & 16.068(015) & \nodata     & DAO  & YALO \\
498.54 & 17.537(017) & \nodata     & \nodata     & \nodata     & \nodata     & \nodata     & PHOT & CTIO 1.5-m \\
498.55 & 17.454(077) & 17.351(019) & 16.647(015) & 16.310(011) & 16.072(015) & \nodata     & DAO  & YALO \\
498.56 & 17.559(022) & 17.332(014) & 16.668(015) & 16.331(015) & 16.097(015) & \nodata     & PHOT & CTIO 1.5-m \\
498.57 & \nodata     & 17.334(014) & 16.670(015) & 16.325(015) & 16.103(015) & \nodata     & PHOT & CTIO 1.5-m \\
499.56 & 17.562(060) & 17.363(017) & 16.615(015) & 16.285(011) & 16.029(015) & \nodata     & DAO  & YALO \\
499.57 & 17.528(079) & 17.375(017) & 16.629(015) & 16.283(011) & 16.015(015) & \nodata     & DAO  & YALO \\
499.66 & 17.607(022) & 17.376(014) & 16.671(015) & 16.299(015) & 16.050(015) & \nodata     & DAO  & CTIO 0.91-m \\
500.54 & 17.574(070) & 17.408(017) & 16.628(015) & 16.263(011) & 15.986(015) & \nodata     & DAO  & YALO \\
500.55 & 17.614(085) & 17.411(017) & 16.608(015) & 16.257(011) & 15.983(015) & \nodata     & DAO  & YALO \\
501.52 & 17.618(103) & 17.450(017) & 16.600(015) & 16.247(011) & 15.984(015) & \nodata     & DAO  & YALO \\
501.53 & 17.720(090) & 17.418(017) & 16.619(015) & 16.253(011) & 15.967(015) & \nodata     & DAO  & YALO \\
501.55 & 17.855(026) & \nodata     & \nodata     & \nodata     & \nodata     & \nodata     & DAO  & CTIO 0.91-m \\
501.57 & 17.811(027) & 17.474(014) & 16.680(015) & 16.285(015) & 16.039(015) & 15.823(015) & DAO  & CTIO 0.91-m \\
503.52 & 17.981(159) & 17.541(017) & 16.660(015) & 16.260(011) & 15.966(015) & \nodata     & DAO  & YALO \\
503.53 & 18.214(167) & 17.563(017) & 16.639(015) & 16.267(011) & 15.974(015) & \nodata     & DAO  & YALO \\
504.53 & \nodata     & 17.623(021) & 16.691(015) & 16.279(011) & 15.988(015) & \nodata     & DAO  & YALO \\
504.54 & \nodata     & 17.620(020) & 16.678(015) & 16.270(011) & 15.970(015) & \nodata     & DAO  & YALO \\
505.52 & \nodata     & 17.717(020) & 16.723(015) & 16.316(011) & 15.988(015) & \nodata     & DAO  & YALO \\
505.53 & 18.352(155) & 17.708(019) & 16.716(015) & 16.294(011) & 15.994(015) & \nodata     & DAO  & YALO \\
506.55 & 18.484(050) & 17.815(014) & 16.845(015) & 16.343(015) & 16.072(015) & 15.855(017) & DAO  & CTIO 0.91-m \\
507.58 & \nodata     & 17.932(024) & 16.851(015) & 16.398(011) & 16.038(015) & \nodata     & DAO  & YALO \\
507.59 & \nodata     & 17.930(024) & 16.877(015) & 16.390(011) & 16.070(015) & \nodata     & DAO  & YALO \\
508.54 & \nodata     & 18.066(018) & 16.922(015) & 16.448(011) & 16.089(015) & \nodata     & DAO  & YALO \\
508.55 & \nodata     & 18.060(018) & 16.907(015) & 16.433(011) & 16.120(015) & \nodata     & DAO  & YALO \\
508.58 & 18.966(041) & 18.062(014) & 16.972(015) & 16.435(015) & 16.130(015) & 15.875(015) & DAO  & CTIO 0.91-m \\
509.55 & 19.041(045) & 18.154(014) & \nodata     & 16.496(015) & 16.173(015) & 15.925(015) & DAO  & CTIO 0.91-m \\
511.56 & \nodata     & 18.437(021) & 17.203(015) & 16.654(011) & 16.232(015) & \nodata     & DAO  & YALO \\
511.57 & \nodata     & 18.452(023) & 17.196(015) & 16.651(011) & 16.244(015) & \nodata     & DAO  & YALO \\
513.59 & \nodata     & 18.674(057) & 17.421(025) & 16.799(013) & 16.337(020) & \nodata     & DAO  & YALO \\
513.60 & \nodata     & 18.580(049) & 17.371(023) & 16.782(013) & 16.337(021) & \nodata     & DAO  & YALO \\
515.59 & \nodata     & 18.801(052) & 17.514(021) & 16.937(012) & 16.484(021) & \nodata     & DAO  & YALO \\
515.60 & \nodata     & 18.942(052) & 17.489(020) & 16.921(012) & 16.452(016) & \nodata     & DAO  & YALO \\
517.53 & \nodata     & 19.035(040) & 17.663(018) & 17.046(012) & 16.573(016) & \nodata     & DAO  & YALO \\
517.54 & \nodata     & 19.073(042) & 17.688(017) & 17.058(011) & 16.568(017) & \nodata     & DAO  & YALO \\
519.53 & \nodata     & 19.173(053) & 17.803(020) & 17.162(012) & 16.629(020) & \nodata     & DAO  & YALO \\
519.54 & \nodata     & 19.214(053) & 17.774(016) & 17.162(011) & 16.639(017) & \nodata     & DAO  & YALO \\
521.58 & \nodata     & 19.202(046) & 17.894(024) & 17.245(013) & 16.684(020) & \nodata     & DAO  & YALO \\
521.59 & \nodata     & 19.254(045) & 17.875(029) & 17.242(014) & 16.770(020) & \nodata     & DAO  & YALO \\

\enddata
\tablecomments{DAO means PSF photometry, while PHOT means aperture photometry}
\end{deluxetable}

\clearpage

\begin{deluxetable} {ccccc}
\tablecolumns{5}
\tablenum{8}
\tablewidth{0pc}
\tablecaption{Temperatures, Angular Radii, and Luminosities for SN~1999ex \label{tal}}
\tablehead{
\colhead{$t-t_0$} & 
\colhead{$T_{BVI}$} & 
\colhead{$\theta_{BVI}$} &
\colhead{$log L(BB)$} &
\colhead{$log L(U\rightarrow z)$} \\
\colhead{(days)} &
\colhead{(K)} &
\colhead{(10$^{11}$cm Mpc$^{-1}$)} &
\colhead{(ergs s$^{-1}$)} &
\colhead{(ergs s$^{-1}$)} }
\startdata
   1.04& 12918(1415) &  23(3) & 41.440 & 41.021 \\
   2.06&  9656(1199) &  36(7) & 41.332 & 41.063 \\
   2.09&  7998(821)  &  48(9) & 41.243 & 41.046 \\
   2.10&  7373(1088) &  60(17)& 41.303 & 41.082 \\
   2.11&  8210(1768) &  49(18)& 41.302 & 41.067 \\
   3.11&  8721(547)  &  54(5) & 41.503 & 41.196 \\
   3.12&  8334(485)  &  57(5) & 41.470 & 41.190 \\
   4.00&  7747(233)  &  83(4) & 41.664 & 41.363 \\
   4.03&  7580(468)  &  91(10)& 41.705 & 41.403 \\
   4.04&  7900(486)  &  83(8) & 41.704 & 41.419 \\
   5.04&  7360(164)  & 117(5) & 41.878 & 41.621 \\
   5.97&  8656(126)  & 105(2) & 42.060 & 41.790 \\
   6.05&  8415(290)  & 116(7) & 42.097 & 41.812 \\
   6.06&  8945(329)  & 106(6) & 42.126 & 41.818 \\
   6.95&  9006(134)  & 114(3) & 42.206 & 41.925 \\
   7.91&  9705(144)  & 110(2) & 42.304 & 42.013 \\
   7.94&  9731(144)  & 111(2) & 42.311 & 42.011 \\
   7.99&  9794(146)  & 112(2) & 42.329 & 42.021 \\
   8.04&  9485(207)  & 122(4) & 42.353 & 42.039 \\
   8.05&  9425(183)  & 126(4) & 42.365 & 42.049 \\
   8.93&  9718(149)  & 125(3) & 42.415 & 42.102 \\
   9.91& 10115(157)  & 127(3) & 42.498 & 42.169 \\
  10.01&  9753(164)  & 137(3) & 42.503 & 42.193 \\
  10.02& 10011(173)  & 133(3) & 42.518 & 42.198 \\
  11.88& 10119(152)  & 144(3) & 42.607 & 42.273 \\
  11.98& 10061(164)  & 147(3) & 42.614 & 42.291 \\
  11.99& 10013(163)  & 148(3) & 42.614 & 42.279 \\
  12.89& 10067(150)  & 150(3) & 42.635 & 42.307 \\
  14.90&  9630(141)  & 169(4) & 42.662 & 42.353 \\
  14.92&  9441(143)  & 177(4) & 42.667 & 42.369 \\
  14.93&  9345(139)  & 180(4) & 42.664 & 42.372 \\
  15.87&  9225(124)  & 185(4) & 42.663 & 42.369 \\
  15.97&  9320(122)  & 183(4) & 42.673 & 42.373 \\
  16.84&  8970(134)  & 199(5) & 42.682 & 42.391 \\
  16.85&  8953(127)  & 198(4) & 42.673 & 42.391 \\
  17.83&  8651(118)  & 212(5) & 42.670 & 42.394 \\
  17.83&  8651(118)  & 212(5) & 42.670 & 42.394 \\
  17.84&  8790(129)  & 206(5) & 42.676 & 42.395 \\
  17.85&  8853(113)  & 202(4) & 42.671 & 42.388 \\
  17.86&  8875(114)  & 201(4) & 42.670 & 42.388 \\
  18.84&  8453(112)  & 221(5) & 42.669 & 42.398 \\
  18.85&  8316(108)  & 227(5) & 42.662 & 42.397 \\
  18.94&  8382(100)  & 221(4) & 42.652 & 42.387 \\
  19.81&  8035(100)  & 240(5) & 42.652 & 42.397 \\
  19.82&  8022(100)  & 242(5) & 42.654 & 42.398 \\
  20.78&  7877(96)   & 248(5) & 42.646 & 42.396 \\
  20.79&  7912(97)   & 247(5) & 42.650 & 42.395 \\
  20.81&  7900(92)   & 244(5) & 42.636 & 42.382 \\
  20.83&  7888(88)   & 241(5) & 42.621 & 42.374 \\
  22.75&  7432(85)   & 270(6) & 42.616 & 42.371 \\
  22.76&  7401(84)   & 271(6) & 42.615 & 42.365 \\
  23.75&  7366(92)   & 270(6) & 42.602 & 42.350 \\
  23.76&  7281(87)   & 277(6) & 42.604 & 42.353 \\
  24.73&  7026(81)   & 289(6) & 42.579 & 42.332 \\
  24.74&  7045(79)   & 288(6) & 42.579 & 42.335 \\
  25.75&  6776(63)   & 293(6) & 42.528 & 42.299 \\
  26.77&  6656(78)   & 305(7) & 42.531 & 42.280 \\
  26.78&  6739(80)   & 294(7) & 42.523 & 42.276 \\
  27.72&  6261(60)   & 329(7) & 42.491 & 42.249 \\
  27.73&  6373(63)   & 316(6) & 42.488 & 42.251 \\
  27.76&  6246(53)   & 323(6) & 42.473 & 42.242 \\
  28.72&  6121(51)   & 327(6) & 42.447 & 42.215 \\
  30.70&  5794(55)   & 346(7) & 42.402 & 42.139 \\
  30.71&  5859(59)   & 338(7) & 42.401 & 42.138 \\
  32.71&  5750(100)  & 328(12)& 42.341 & 42.061 \\
  32.72&  5902(100)  & 316(11)& 42.353 & 42.075 \\
  34.69&  5889(100)  & 297(10)& 42.296 & 42.005 \\
  34.70&  5787(84)   & 309(9) & 42.301 & 42.003 \\
  36.60&  5580(70)   & 312(8) & 42.244 & 41.945 \\
  36.61&  5517(70)   & 319(9) & 42.244 & 41.939 \\
  38.58&  5413(81)   & 319(11)& 42.212 & 41.897 \\
  38.59&  5574(75)   & 302(9) & 42.216 & 41.897 \\
  40.61&  5272(76)   & 325(11)& 42.182 & 41.867 \\
  40.62&  5381(82)   & 302(10)& 42.154 & 41.858 \\
\enddata
\tablecomments{Adopted values: $E(B-V)_{Gal}$=0.02, $E(B-V)_{host}$=0.28, $D$=51.16 Mpc, $t_0$=JD 2,451,480.5}
\end{deluxetable}

\end{document}